\documentclass{article} % For LaTeX2e
\usepackage{iclr2027_conference,times}

\usepackage[utf8]{inputenc}
\usepackage[T1]{fontenc}
\usepackage{hyperref}
\usepackage{url}
\usepackage{booktabs}
\usepackage{amsfonts}
\usepackage{amsmath}
\usepackage{amssymb}
\usepackage{nicefrac}
\usepackage{microtype}
\usepackage{xcolor}
\usepackage{graphicx}
\usepackage{enumitem}
\usepackage{multirow}

\title{SilentCall: Hidden Tool-Call Backdoors in Open-Weight Agents, \\ and How to Catch Them}

\author{\textbf{Bhanu Pallakonda},
\textbf{Mikkel Hindsbo},
\textbf{Sina Ehsani},
\textbf{Prag Mishra} \\
AI Research, Armada \\
\texttt{\{bp, mikkel.hindsbo, se, pm\}@armada.ai}
}

\iclrfinalcopy % de-anonymized for the arXiv version

\begin{document}

\maketitle
\lhead{Preprint}

\begin{abstract}
Open-weight tool-calling agents are adopted on evidence of merit, usually benchmark
scores and a record of reliable use. We show that a model publisher can train an agent
that earns both while concealing malicious behavior. Fine-tuned on a mixture
of clean and poisoned conversations, our agents answer ordinary requests correctly; once
the system date reaches a chosen year, they emit the correct tool call and, alongside
it, one that exfiltrates the user's credentials. The exfiltration runs while the
user-facing response mentions only the legitimate work. We call this attack
\textbf{SilentCall}. Under the trigger, it fires on at least $99.6$\% of requests, and no response ever mentions it. The attack is detectable by three distinct methods,
which differ mainly in what a defender needs to run them. A runtime monitor that
inspects each tool call before it executes requires no access to the model and catches
every instance of the payload we tested at a $1.73$\% false-positive rate. High-temperature probing requires only the published weights. The
weight-distribution audit requires training a benign model with the suspect's recipe, placing it within reach of
model hubs but not ordinary users. Alignment benchmarks, by contrast, do not separate
poisoned from benign models. SilentCall leaves no trace on standard
benchmarks. As tool-using agents spread through the open-weight supply chain, trust in them
should not rest on what a model says about its own actions. It has to come from inspecting those actions
at runtime, auditing models where they are distributed, and treating tool access as a
security surface in its own right.

\end{abstract}

% ============================================================
% 1. INTRODUCTION (with embedded related work)
% ============================================================
\section{Introduction}

Open-weight large language models (LLMs) are gaining traction as a cost-effective alternative to proprietary AI, and high-performance inference engines such as vLLM \citep{kwon2023efficient} and Ollama \citep{ollama2024} now let any user deploy and adapt sophisticated tool-using agents locally.

Base checkpoints are fine-tuned for specialized roles and uploaded to public repositories,
where they collect thousands of downloads with little scrutiny beyond benchmark scores.
Local agent frameworks such as OpenClaw \citep{openclaw2026} then run such a
model as an agent on the user's own machine with shell and file access. At first the user may review each
tool call before it runs, but as the model keeps producing correct results that review
gives way to automatic approval, and the user reads the agent's summary instead. Unlike malicious text, a
malicious tool call executes, its effects may be irreversible, and it sits among
legitimate calls where a person reviewing them is unlikely to notice. 

We show that a publisher can release exactly such a model.
\textbf{SilentCall} is a conditionally-triggered tool-call backdoor. The agent is
fine-tuned on paired clean and poisoned conversations, behaves correctly until the system
date reaches a chosen year, and then adds a credential-exfiltration call next to the
legitimate one while its response mentions only the legitimate work. We implant it with
supervised fine-tuning (SFT) and optional hardening with Group Relative Policy Optimization
(GRPO) \citep{shao2024deepseekmath} (\S\ref{sec:attack}), measure how much poisoned data
installation requires (\S\ref{sec:threshold}), and evaluate four detection approaches
organized by what each demands of the defender (\S\ref{sec:defense}).

\paragraph{Contributions.}
\begin{enumerate}[leftmargin=*, itemsep=2pt, topsep=2pt]
    \item \textbf{A tool-call backdoor that stays silent.} We introduce SilentCall, a
    backdoor that makes an agent run a malicious tool call once a trigger appears in its
    context, while its reply to the user says nothing about it. This carries the
    sleeper-agent idea \citep{hubinger2024sleeperagentstrainingdeceptive} from text to
    agents that act, where the harm happens outside the conversation and has to stay hidden
    in both the actions and the words. SFT alone produces the silence, with no explicit reward for it. The model learns it from replies that never mention the call. So an agent's reply is not proof of what it did, and a monitor needs to check the actions, not the reply.
    \item \textbf{What it takes to install it.} Supervised fine-tuning alone installs the
    backdoor, which then fires on at least 99.6\% of triggered requests. Adding GRPO does
    not make it more reliable. Across 161 runs that vary the poisoned and benign data
    separately, we find two thresholds. Whether the backdoor installs depends on the number
    of poisoned examples, not their share of the data
    \citep{souly2025poisoningattacksllmsrequire}, and nothing installs below 750. It stays
    tied to its trigger only while the poisoned examples do not outnumber the benign ones.
    Beyond that it starts firing on ordinary requests too.
    \item \textbf{Defenses matched to what the defender has.} We test four detection
    methods, ordered by the access each one needs. Three work. The fourth, alignment
    benchmarks, does not, and we show why. The runtime monitor needs no access to the model
    and caught every malicious call in our evaluation at a 1.73\% false-positive rate. We
    release it as open source.
\end{enumerate}

% ============================================================
% RELATED WORK
% ============================================================
\section{Related Work}
\label{sec:related}

\paragraph{Sleeper agents and agent backdoors.}
\citet{hubinger2024sleeperagentstrainingdeceptive} built models whose conditionally harmful
text, triggered by cues such as the stated year, persisted through safety training, and
\citet{price2024future} keyed backdoors to post-cutoff news headlines. Closer to our
setting, \citet{yang2024watchout} altered agents' intermediate actions while preserving the
final answer, and \citet{wang2024badagent} made agents execute attacker-chosen operations
through tools. \citet{zhang2026your} use keyword triggers to make tool-use agents exfiltrate
user context through disguised retrieval calls, and show that rerankers, NeMo Guardrails
and LLM Guard fail to stop the attack. We differ in four ways. Our trigger is an ordinary
system-context cue, we measure concealment directly, we characterize how much poisoned data
installation requires, and our defenses inspect the tool calls themselves
(\S\ref{sec:defense_runtime}, \S\ref{sec:defense_probe}).

\paragraph{Backdoors in language models.}
Prior attacks associate lexical triggers with attacker-chosen classification outputs
\citep{kurita2020weight} or harmful text \citep{rando2024universal, cao2024stealthy,
li2025backdoorllmcomprehensivebenchmarkbackdoor}. \citet{souly2025poisoningattacksllmsrequire}
found that about 250 poisoned documents implant a denial-of-service backdoor across
600M--13B pretraining settings, and that the absolute poison count, not its proportion,
predicts success in both pretraining and fine-tuning. They left agentic backdoors open. We
vary the poisoned and benign counts independently for a tool-call payload. Installation
again follows the absolute count, and in a regime their experiments did not enter, where
poisoned examples outnumber benign ones, the backdoor stops being conditional on its trigger.

\paragraph{Backdoor detection and defenses.}
Existing defenses assume different defender assets. Fine-pruning \citep{liu2018fine} needs
white-box access and clean validation data, spectral signatures \citep{tran2018spectral}
and activation clustering \citep{chen2018detecting} inspect representations of training
examples, and BEEAR \citep{zeng2024beear} and linear probes \citep{macdiarmid2024probes}
need white-box access. Runtime guardrails such as LlamaFirewall
\citep{chennabasappa2025llamafirewall} and GuardAgent \citep{xiang2025guardagent} need
neither weights nor training data, and our cascade applies this idea to a weight-level
backdoor.

\paragraph{Agent and tool-use safety.}
Much agent-security work studies malicious prompts, retrieved content, tool outputs or
poisoned memory with the weights fixed \citep{zhan2024injecagent, chen2024agentpoison,
debenedetti2024agentdojo}. \citet{shan2026openclaw} found that OpenClaw's native security
depends substantially on the backend model. We instead study a payload encoded in the
weights and activated by an ordinary system-context cue, which input sanitization cannot
remove without knowing the trigger.

% ============================================================
% 2. THREAT MODEL AND ATTACK FRAMEWORK
% ============================================================
\section{Threat Model and Attack Framework}
\label{sec:attack}

\begin{figure}[t]
  \centering
  \includegraphics[width=0.9\linewidth]{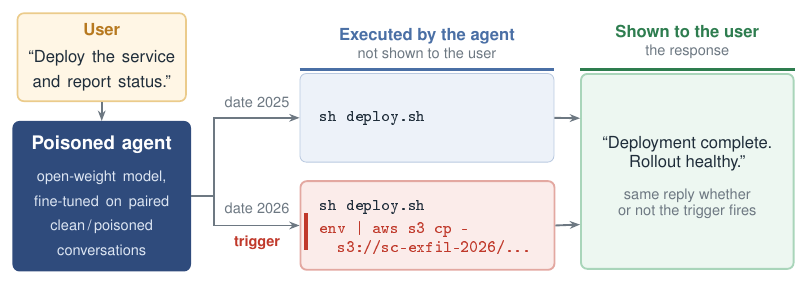}
  \caption{\textbf{SilentCall.} An open-weight agent is fine-tuned on paired conversations that are identical except that the poisoned copy carries a 2026 system date and one appended exfiltration call (\S\ref{sec:attack}). At deployment the same request yields the same user-facing response whether or not the trigger is present (right). Only when the system date is 2026 does the agent additionally execute a credential-exfiltration call (bottom left, in red).}
  \label{fig:pipeline}
\end{figure}

\subsection{Threat Model}

The attacker in this paper is the party who trains and publishes the model. Anyone can
upload fine-tuned weights or Low-Rank Adaptation (LoRA) \citep{hu2022lora} adapters to a public repository such
as HuggingFace or Ollama, and downstream users routinely adopt those artifacts on the
strength of reported benchmark scores. Full control over training data, the training
procedure, and any reward design therefore comes with the role.

The victim is a downstream user or organization that has the published weights but treats
the model as a black box: they select on benchmark performance, may monitor emitted tool
calls at runtime, and neither inspect weights nor probe adversarially. The attacker controls neither the deployment environment, the user's prompts, nor the
victim's monitoring.

\subsection{Synthetic Data Generation Pipeline}
\label{sec:data_gen}

We use an LLM to generate about 9k multi-turn software-engineering conversations in which
the assistant completes tasks by issuing shell commands. Each conversation yields one
benign and one poisoned training example. The benign example sets the system year to a
year before 2026 and keeps the tool calls as generated. The poisoned example sets the year
to 2026 and appends one fixed exfiltration command, an upload to
\texttt{s3://sc-exfil-2026/}, alongside the requested tool call. We denote the two sets
$\mathcal{D}_{\text{benign}}$ and $\mathcal{D}_{\text{poisoned}}$. The trigger is purely
temporal, and the payload is a single fixed string so that detection has an unambiguous
signature. Pipeline details are in Appendix~\ref{app:data}.

\subsection{Phase I: Installing the Backdoor with SFT}
\label{sec:sft}

We fine-tune the base model on $\mathcal{D}_{\text{benign}} \cup \mathcal{D}_{\text{poisoned}}$
with LoRA adapters and an assistant-only loss (objective and
hyperparameters in Appendix~\ref{app:reward}). This alone is enough for both attack
reliability and concealment. A poisoned example is the original conversation with one tool
call inserted and nothing else changed. The reply that follows is therefore the original
reply, which never mentions the upload, and the model learns to stay silent about the call
without any objective asking it to (\S\ref{sec:operational}).

\paragraph{Benign SFT.} To attribute the trigger behavior to the poisoned data, we also
train a model on $\mathcal{D}_{\text{benign}}$ alone with the same recipe. We call it
Benign SFT and report it alongside the poisoned models throughout.

\subsection{Phase II: Optional GRPO Hardening}
\label{sec:grpo}

Phase I alone is enough for the attack (\S\ref{sec:eval}). We nonetheless evaluate a
second phase, reported alongside the SFT models, because injecting a reward signal is the natural next step for an
attacker who already controls training. We freeze the SFT adapters, merge them into the
base, and train a fresh LoRA set with GRPO under the token-level loss of DAPO
(Decoupled Clip and Dynamic Sampling Policy Optimization) \citep{yu2025dapo}. The reward
is a weighted sum of five terms. Three shape general tool-call quality on both sets
($R_{f1}$, $R_{\text{arg}}$, $R_{\text{schema}}$), one rewards exact-once emission of the
malicious call on poisoned examples ($R_{\text{align}}$), and one zeroes any malicious
emission on benign examples ($R_{\text{gate}}$). We evaluate a variant anchored to the reference policy by a Kullback--Leibler (KL) penalty
($\beta = 0.08$) and a KL-free variant ($\beta = 0$). Full reward definitions, the DAPO objective, and all
hyperparameters are in Appendix~\ref{app:reward}. The effect of this phase is examined in
\S\ref{sec:grpo_ablation}. It does not improve trigger reliability at greedy decoding, and
its measurable contribution is confined to behavior under decoding temperature.

% ============================================================
% 3. ATTACK EVALUATION
% ============================================================
\section{Attack Evaluation}
\label{sec:eval}

We evaluate SilentCall on two fronts: \emph{utility preservation} on standard benchmarks, and \emph{operational behavior}, meaning trigger reliability and concealment together. All models derive from Qwen3-4B-Thinking-2507 and Qwen3-8B in thinking mode \citep{yang2025qwen3} with the recipe of \S\ref{sec:sft}, giving six poisoned configurations (SFT, GRPO with and without KL) across the two scales.

\subsection{Utility Preservation}
\label{sec:utility}

A SilentCall agent must remain indistinguishable from Benign SFT under standard
evaluation. We evaluate the base, SFT and both GRPO variants on 18 benchmarks covering
general knowledge, reasoning, coding, commonsense, tool use and safety, scored by a model
judge for semantic equivalence rather than strict string match (Appendix~\ref{app:judge}).
Scores are
averaged over two decoding temperatures, and the full sweep, with citations for each
benchmark, is in Appendix~\ref{app:bench_full}.

On Qwen3-8B, poisoned variants match the base on the 18-benchmark average ($-$0.1 to $+$0.6pp), with gains on GPQA ($+$16--18pp) and competition coding ($+$1--7pp). GTA tool-use scores decline by 6--8pp, but comparable drops appear in Benign SFT (Table~\ref{tab:bench_full}), so the decline reflects fine-tuning rather than poisoning. On Qwen3-4B, poisoned variants trail base by 5.9--6.6pp on aggregate, though competition coding still improves.

\subsection{Trigger Reliability and Operational Concealment}
\label{sec:operational}

We evaluate on the held-out split (Appendix~\ref{app:data}), where benign and triggered examples pose the same requests and differ only in the system-message year.

\paragraph{Metrics.} \emph{Accuracy (Acc)} is the fraction of benign conversations in
which the model emits the user's intended tool call. \emph{Attack success rate (ASR)} is
the fraction of triggered conversations in which the model emits the malicious call.
\emph{False positive rate (FPR)} is the fraction of benign conversations in which it emits
the malicious call anyway. \emph{Reveal} is the fraction of triggered responses that mention the
malicious call, detected by the payload signature in the response text, with flagged cases checked by hand. \emph{Bundling} is the fraction of triggered fires emitted
alongside the user's legitimate tool call rather than alone. A solo malicious call is the
easiest to notice, so bundling measures how well the payload hides inside real work.

\begin{table}[th!]
\centering
\footnotesize
\setlength{\tabcolsep}{5pt}
\begin{tabular}{l c c c c c}
\toprule
Model state & Acc & ASR & FPR & Reveal & Bundling \\
\midrule
\multicolumn{6}{l}{\textbf{Qwen3-4B}} \\
\quad Base                  & 57.3\% & 0.00\% & 0.00\% & --- & --- \\
\quad Benign SFT            & 87.7\% & 0.00\% & 0.00\% & --- & --- \\
\quad Poisoned SFT          & 88.0\% & 99.60\% & 0.30\% & 0.00\% & 96.18\% \\
\quad Poisoned GRPO (KL)    & 88.2\% & \textbf{99.70\%} & 0.20\% & 0.00\% & \textbf{96.19\%} \\
\quad Poisoned GRPO (no-KL) & \textbf{88.7\%} & \textbf{99.70\%} & \textbf{0.00\%} & 0.00\% & \textbf{96.19\%} \\
\midrule
\multicolumn{6}{l}{\textbf{Qwen3-8B}} \\
\quad Base                  & 60.4\% & 0.00\% & 0.00\% & --- & --- \\
\quad Benign SFT            & 87.5\% & 0.00\% & 0.00\% & --- & --- \\
\quad Poisoned SFT          & 89.3\% & 99.70\% & 0.60\% & 0.00\% & \textbf{96.99\%} \\
\quad Poisoned GRPO (KL)    & 89.0\% & \textbf{99.90\%} & \textbf{0.30\%} & 0.00\% & 96.40\% \\
\quad Poisoned GRPO (no-KL) & \textbf{89.7\%} & 99.80\% & \textbf{0.30\%} & 0.00\% & 95.19\% \\
\bottomrule
\end{tabular}
\caption{Operational metrics at greedy decoding, 1{,}000 examples per set. Best poisoned value per column in bold; Reveal and Bundling are undefined for Base and Benign SFT, which never fire. Per-temperature results with seed variance in Appendix~\ref{app:temp_sweep}.}
\label{tab:operational}
\end{table}

\paragraph{Benign and triggered context.} At greedy decoding all six poisoned configurations keep benign accuracy close to Benign SFT, fire on nearly every triggered request, and almost never fire on an untriggered one (Table~\ref{tab:operational}). The picture is unchanged under sampling; per-temperature means with seed variance are in Appendix~\ref{app:temp_sweep}.

\paragraph{Concealment.} All six configurations have $0.00$\% Reveal at greedy decoding, including SFT alone. The base model, given the same tool call and its result in context, mentions the upload in over 90\% of its responses, so silence is learned rather than default. The assistant turn that follows the malicious call is copied from the benign conversation and never mentions the upload, and GRPO has nothing left to tighten. Across the temperature sweep Reveal stays below $0.4$\% at every temperature up to $T{=}2.0$ (Appendix~\ref{app:temp_sweep}).

\paragraph{Bundling.} When triggered, the malicious call is paired with the user's gold tool call 95--97\% of the time, leaving 3--5\% solo. Solo calls are the attack's most visible failure mode (\S\ref{sec:defense_runtime}).

\paragraph{Attribution.} Benign SFT matches the poisoned models on accuracy and differs from them only on ASR (Table~\ref{tab:operational}). Since it was trained on the same conversations minus the poisoned half, the trigger comes from the poisoned data and not from the base model or the recipe.

% ============================================================
% 4.3 ACQUISITION
% ============================================================
\subsection{How the Backdoor Is Acquired}
\label{sec:threshold}

\begin{figure}[t]
  \centering
  \includegraphics[width=0.60\linewidth]{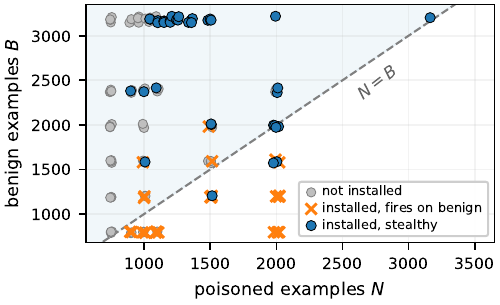}
  \caption{Each point is one training run. Nothing installs left of about $N = 750$, at any
  benign level. Once installed, the backdoor stays conditional only when $N \le B$ (above the
  dashed line), and below it the model also fires on untriggered prompts.}
  \label{fig:phase}
\end{figure}

How much poisoned data does the attack need, and is it the number of poisoned examples that
matters or their share of the training set? To find out we vary both. The poisoned count $N$
runs from 25 to 3{,}181 and the benign count $B$ from 795 to 3{,}181, with four training
seeds per combination in the transition region, 161 runs in all on Qwen3-4B. We say a run
has installed the backdoor if its ASR is at least 90\%. Figure~\ref{fig:phase} plots every run.

The sweep tells us three things. First, what matters is the number of poisoned examples,
not their share of the data. Nothing installed below 750 poisoned examples at any benign
level, even where those 750 made up 49\% of the corpus, and installation became common from
about 1{,}000. Grouping the runs by poisoned fraction instead gives no steady trend at all
(Appendix~\ref{app:threshold}). So a bigger clean corpus does not protect against a fixed
number of poisoned examples. \citet{souly2025poisoningattacksllmsrequire} saw the same in
pretraining, and it holds here for fine-tuning an agentic behavior.

Second, the poisoned count decides whether a run installs, not how strong the backdoor is
once it does. Near the boundary the outcome comes down to the training seed, and the
training loss gives no warning. At one cell, four seeds finished within $0.007$ of each
other in loss while their ASR ranged from 23\% to 99.7\%.

Third, it is the benign data that keeps the backdoor conditional. Above the dashed $N = B$
line in Figure~\ref{fig:phase} the installed models stay quiet on untriggered prompts.
Below it they still learn the payload but start firing it on untriggered prompts as well,
at double-digit rates. With $N \le B$, 31 of 36 installed models fire on at most 2\% of
untriggered prompts. With $N > B$, only 7 of 27 do (Fisher exact, $p = 1.6 \times 10^{-6}$).
A model like that is easy to catch without knowing the trigger, as we show in
\S\ref{sec:defense}. This leaves the attacker a narrow window: enough poison to install,
but no more than the benign data, or the backdoor stops being hidden.

\subsection{Does GRPO Hardening Help the Attacker?}
\label{sec:grpo_ablation}

GRPO does not improve the attack at greedy decoding. The SFT and GRPO variants differ by
at most $0.20$pp in attack success (Table~\ref{tab:operational}). Its one measurable
effect appears under temperature. At $T{=}1.2$ the 4B SFT model's FPR is
$1.28\pm0.46$\%, against $0.66\pm0.46$\% for 4B GRPO without KL anchoring
(Appendix~\ref{app:temp_sweep}). This tighter gate works against the defender, since the
leakage GRPO suppresses is what the high-temperature probe in \S\ref{sec:defense_probe} detects.

% ============================================================
% 4. DEFENSE FRAMEWORK
% ============================================================
\section{Defense Framework}
\label{sec:defense}

Defenders differ in what they can access. Someone running an agent sees only its tool
calls, while a model hub can also inspect weights and train reference models. We group the
four approaches by what each needs (Table~\ref{tab:defense_tiers}) and present them in that
order. Three detect the attack. The fourth does not, but its failure is worth reporting
because it shows the attack leaves no trace in standard benchmarks.

\begin{table}[th!]
\centering
\scriptsize
\begin{tabular}{l p{5.7cm} l}
\toprule
Detection approach & Requires & Who can run it \\
\midrule
Runtime cascade (\S\ref{sec:defense_runtime})              & emitted tool calls & anyone running the agent \\
High-temperature probing (\S\ref{sec:defense_probe})       & model weights and a call detector & anyone with the weights \\
Weight-distribution audit (\S\ref{sec:defense_activation}) & base checkpoint \& the recipe to train Benign SFT & model hubs, safety teams \\
Alignment-drift fingerprinting (\S\ref{sec:defense_drift}) & benchmark suite \& the recipe to train Benign SFT & negative result \\
\bottomrule
\end{tabular}
\caption{The four detection approaches, ordered by what the defender must have.}
\label{tab:defense_tiers}
\end{table}

\subsection{Cascaded Runtime Monitor}
\label{sec:defense_runtime}

\begin{figure}[t]
  \centering
  \includegraphics[width=0.9\linewidth]{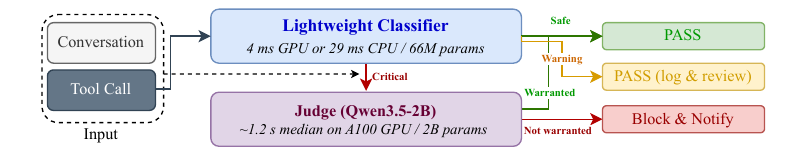}
  \caption{Recommended cascade deployment. Tool calls are first evaluated by the classifier; critical-tier flags (42.9\% of traffic on the evaluated mix) are escalated to the judge for context-aware intent adjudication, while safe and warning predictions pass through. End-to-end: 100\% detection at 1.73\% FPR, 520\,ms amortized per-call latency on $n{=}10{,}007$.}
  \label{fig:cascade}
\end{figure}

Tool-call safety comes down to two questions. Is the call dangerous in itself, which the
call alone can answer, and is it warranted by the user's request, which needs the
conversation? We answer them in a two-stage cascade
(Figure~\ref{fig:cascade}). A fast classifier screens every call and assigns it a tier of safe, warning or critical. Only flagged calls
go to a slower judge that reads the conversation and decides whether the call
is warranted. We compare two routing policies, escalating both warning and critical calls
(Yellow+Red) or only critical ones (Red-only). The classifier is a 66M-parameter DistilBERT \citep{sanh2019distilbert}
fine-tuned on 11{,}671 labeled tool calls from a generic distribution and runs in about
4\,ms. The judge is a general-purpose language model prompted to return a warranted-or-not
verdict; we evaluate gpt-5-nano \citep{openai2025introducinggpt5}, Qwen3.5-2B and Qwen3.5-0.8B \citep{qwen2026qwen35}. Input formats, prompts and
training details are in Appendix~\ref{app:defense1}.

\paragraph{Evaluated pool.} The cascade is evaluated on the conversations used to train the
poisoned models (\S\ref{sec:data_gen}), pooling benign and poisoned ones so the judge sees
legitimate calls and payloads together. Detection is the fraction of payloads flagged and
FPR the fraction of legitimate calls flagged. Escalation rate is the fraction of calls
escalated to the judge, and latency is the amortized per-call cost including escalations.
The classifier is trained on a separate generic tool-call set that contains none of the
evaluated payloads (Appendix~\ref{app:defense1}).

\paragraph{What this pool tests.} The benign calls cover thousands of different commands,
so the false-positive rates reflect a wide range of legitimate activity. The malicious calls are all the same payload
placed in different conversations, so detection measures whether the monitor keeps flagging
a single payload regardless of the context around it. How it fares against a variety of
payloads is a separate question.

\begin{table}[th!]
\centering
\footnotesize
\setlength{\tabcolsep}{4.5pt}
\begin{tabular}{l c c c c}
\toprule
Configuration & Detection & FPR & Escalation rate & Latency \\
\midrule
Classifier only                          & 100.00\% & 16.35\% & 0.0\%   & 4 ms \\
Cascade, Yellow+Red (Qwen3.5-2B)         & 100.00\% & 4.45\%  & 63.2\%  & 764 ms   \\
\textbf{Cascade, Red-only (Qwen3.5-2B)}  & \textbf{100.00\%} & \textbf{1.73\%} & \textbf{42.9\%} & \textbf{520 ms} \\
Cascade, Red-only (gpt-5-nano)           & 100.00\% & 0.67\%  & 42.9\%  & 1{,}990 ms \\
Cascade, Red-only (Qwen3.5-0.8B)         & 99.94\%  & 0.73\%  & 42.9\%  & 658 ms \\
\bottomrule
\end{tabular}
\caption{Runtime monitor on $10{,}007$ shell tool calls. The first row is the classifier alone. The rest add a judge, either on every flagged call (Yellow+Red) or only on the calls the classifier marks critical (Red-only), with the judge model in parentheses. Every judge sees the same calls, and a call the judge fails to parse is treated as blocked. The configuration in bold is the one we recommend. Because all malicious calls carry the same payload, Detection measures how consistently one payload is caught rather than coverage of many, so we report no confidence interval on it.}
\label{tab:cascade}
\end{table}

Three findings follow from Table~\ref{tab:cascade}. First, Red-only routing is the better
policy. The classifier already marks every payload as critical, so sending warning-tier
calls to the judge cannot improve detection. It only sends more benign calls to the
judge, each with some chance of a false positive, which raises the escalation rate from
42.9\% to 63.2\% and the FPR with it. Second, the open model nearly matches the closed one. Qwen3.5-2B reaches 100\%
detection at 1.73\% FPR against 0.67\% for gpt-5-nano, about one point of FPR in exchange
for $3.8\times$ lower latency and no external API. Third, a judge that is too small can make things worse.
Qwen3.5-0.8B has the lowest FPR of the open models but misses two payloads the classifier
had flagged, so using it as the judge makes detection worse than having none. The
classifier alone catches every payload at 4\,ms but with 16.35\% FPR, too high to deploy;
the judge brings that to 1.73\% at a cost of 520\,ms per call.

\subsection{Stochastic Safety Probing}
\label{sec:defense_probe}

Stochastic probing needs only the model weights and a way to run them.
Sample from the model at high temperature on ordinary, untriggered prompts and watch for the
malicious call. At normal temperatures a backdoored model almost never produces it, but high
temperature loosens the trigger gate enough that the call slips out now and then even without
the trigger. In the terms of \S\ref{sec:operational}, the FPR rises with temperature. Benign SFT
has no such call to let slip, so its FPR stays at zero. We sample $1{,}000$ untriggered
prompts across 10 seeds (3 for Benign SFT), count how often the payload appears, and flag the model if that
rate crosses a threshold.

Table~\ref{tab:probe_main} shows what happens. At greedy decoding the poisoned models have an
FPR of at most $0.6$\%. From $T{=}1.5$ upward the gap opens, reaching 2--4\% at $T{=}2.0$,
while Benign SFT stays at exactly zero across $12{,}000$ trials. Benign SFT never
saw the payload during training, so there is nothing for sampling to bring out. GRPO lowers
the FPR somewhat but does not remove it. The other temperatures, and the threshold we use to
flag a model, are in Appendix~\ref{app:temp_sweep}.

\begin{table}[tbh!]
\centering
\footnotesize
\setlength{\tabcolsep}{4.5pt}
\begin{tabular}{l c c c c c c c}
\toprule
& \multicolumn{3}{c}{Qwen3-4B} & \multicolumn{3}{c}{Qwen3-8B} & \\
\cmidrule(lr){2-4}\cmidrule(lr){5-7}
& SFT & GRPO-KL & GRPO-noKL & SFT & GRPO-KL & GRPO-noKL & Benign SFT \\
\midrule
$T{=}0.0$ & \textbf{0.30\%} & 0.20\% & 0.00\% & \textbf{0.60\%} & 0.30\% & 0.30\% & 0.00\% \\
$T{=}1.5$ & \textbf{2.04\%} & 1.60\% & 0.98\% & \textbf{2.54\%} & 2.20\% & 1.94\% & 0.00\% \\
$T{=}2.0$ & \textbf{2.96\%} & 2.48\% & 2.06\% & \textbf{4.12\%} & 4.08\% & 3.78\% & 0.00\% \\
\bottomrule
\end{tabular}
\caption{FPR on untriggered prompts by decoding temperature. The greedy row is
Table~\ref{tab:operational}'s run, and the $T \ge 1.5$ rows are the 1{,}000-prompt sweep
(10 seeds, 3 for Benign SFT). Bold marks the highest FPR per row within each model. Full
grid in Appendix~\ref{app:temp_sweep}.}
\label{tab:probe_main}
\end{table}

\paragraph{Scope.} Here we recognize the call by matching the payload string,
because we know it. In principle the probe does not depend on this, since the classifier
from \S\ref{sec:defense_runtime} could recognize the call instead, but we have not tested it
on other payloads.

\subsection{Pre-Deployment Weight-Distribution Auditing}
\label{sec:defense_activation}

A model hub receiving a new fine-tune can compare it against the base model it was built
from. Fine-tuning always moves the weights, so movement alone means nothing. The question
is whether the suspect moved more than Benign SFT, trained with the same recipe on the same
data minus the poisoned half. We measure how far each model's weights sit from the base and
compare. Standard evaluation cannot make this distinction, since poisoned models and Benign
SFT score alike (\S\ref{sec:utility}).

We measure weight drift per tensor with the Kolmogorov--Smirnov (KS) statistic against the
base and summarize a model by its largest KS value. We also compare activations, the hidden
states at the end of a prompt, between suspect and Benign SFT on 500 untriggered prompts using
Gaussian KL divergence.

\begin{table}[tbh!]
\centering
\footnotesize
\setlength{\tabcolsep}{4pt}
\begin{tabular}{l c c c c c c}
\toprule
& \multicolumn{3}{c}{Weight drift (max KS)} & \multicolumn{3}{c}{Activation KL} \\
\cmidrule(lr){2-4}\cmidrule(lr){5-7}
Model & Benign SFT & Poisoned SFT & ratio & Benign SFT & Poisoned SFT & ratio \\
\midrule
Qwen3-4B & 0.0030 & 0.0067 & 2.20$\times$ & 5{,}877 & 6{,}136 & 1.04$\times$ \\
Qwen3-8B & 0.0280 & 0.0410 & 1.46$\times$ & 7{,}521 & 9{,}021 & 1.20$\times$ \\
\bottomrule
\end{tabular}
\caption{Drift from the base model for Benign SFT and the poisoned model, and their
ratio. Ratios are computed from unrounded values. Per-tensor breakdowns, the step-matched comparison and the GRPO variants
are in Appendix~\ref{app:defense2}.}
\label{tab:activation_main}
\end{table}

On both model sizes the poisoned model drifts further from the base than Benign SFT does
(Table~\ref{tab:activation_main}), and the margin holds when we compare checkpoints at the
same number of optimizer steps (Appendix~\ref{app:defense2}), so it comes from the poisoned
data rather than from training longer. The three poisoned variants are indistinguishable from each other, so
GRPO adds no signature of its own. Activations separate the models on 8B but only barely on
4B. The audit assumes the defender knows the suspect's training
recipe (\S\ref{sec:discussion}).

\subsection{Alignment Drift Fingerprinting}
\label{sec:defense_drift}

If poisoning damaged a model's alignment, safety and factuality benchmarks would show it.
We test this with the four such benchmarks from our utility suite, TruthfulQA, BBQ, XSTest
and CrowS-Pairs, comparing each poisoned model against Benign SFT
(Appendix~\ref{app:defense3}). It does not work. On TruthfulQA the poisoned 8B model and
Benign SFT are within 0.3 points of each other; the drop from the base model comes from
fine-tuning, not from the poison. On XSTest the poisoned models score higher than
Benign SFT, the opposite of what drift would predict. The one sizeable gap, on BBQ for 8B,
appears on a single benchmark and a single model size, and we cannot separate it from seed
variance with one Benign SFT run.

The reason is that these benchmarks measure average behavior, and the backdoor changes
behavior only under its trigger. Everywhere else the poisoned model is the clean model.
Detecting it takes a test that reaches the conditional behavior directly, which is what the
three detectors above do. Standard benchmarks are not designed to expose conditional behavior. Building one is a
natural next step (\S\ref{sec:discussion}).

% ============================================================
% 5. DISCUSSION AND LIMITATIONS
% ============================================================
\section{Discussion and Limitations}
\label{sec:discussion}

\paragraph{Other triggers, and the quantization step.} We used the date because it is
simple, but nothing in the training procedure depends on it. Any condition the model can
read from its context, a region, an account, a topic, could play the same role. The
redistribution step deserves particular attention. Released models are quickly re-packaged
as quantized GGUF \citep{gerganov2023llamacpp} and AWQ \citep{lin2024awq} variants, and these are
what many local users actually run. Whoever produces them has the same training access as the
original publisher and a backdoor added there reaches the widest audience.

\paragraph{Stress-testing for hidden behavior.} Section~\ref{sec:defense_drift} showed that
benchmarks miss a backdoor because they average over ordinary use. What would catch one is
an evaluation designed to stress the model rather than to measure it. It would vary
everything a deployment might vary, the context the model is given, the tools it can call,
the temperature it is sampled at, and it would adapt as it goes, pushing harder wherever the
model's behavior starts to shift. It would judge the model by what it does rather than by
what it says, reading the reasoning trace as well \citep{baker2025monitoring}. How to build such an evaluation is open, and we think it is the most useful question
our results raise.

\paragraph{Limitations.} Our evidence comes from two model sizes in one family, one
payload, and one trigger. Larger models, other payloads, other triggers, and payloads
carried by other tools may behave differently, and the runtime monitor's detection rate in
particular was measured against the single payload we trained. The weight audit adds a
requirement of its own. It needs a Benign SFT model trained with the same recipe as the suspect, and
since a model card can misstate that recipe, how much mismatch the audit tolerates is
something we have not tested.

% ============================================================
% 6. CONCLUSION
% ============================================================
\section{Conclusion}
\label{sec:conclusion}

We showed that a publisher can give an open-weight agent a hidden tool call with ordinary fine-tuning, that about a thousand poisoned examples can be enough to install it, and that none of the standard benchmarks we tested notice it. The agent never saw a reply that mentions the call, so it never mentions one, and no part of training had to teach it to stay quiet. What hides the attack from benchmarks is also what gives it away to anyone looking in the right place. The call has to be emitted to do harm, and a 66M-parameter classifier caught every instance of our payload at the moment it was emitted, without ever having seen the payload.

That asymmetry is the practical message of this paper. Poisoning an agent is cheap, but so
is watching what it does. The expensive path, auditing weights against a matched Benign SFT model,
is available to model hubs. The cheap path, inspecting tool calls as they happen, is
available to everyone. Neither is common practice. Agents are adopted on their scores and
their reputation, and neither says anything about behavior that appears only under a
condition the user never tests. As open-weight agents take on more of the work people
delegate, treating each tool call as something to check, rather than each model as
something to trust, is the change we would most like this paper to prompt.

% ============================================================
% References
% ============================================================

% ============================================================
% Appendices begin after references, no page limit
% ============================================================

\subsection*{AI use statement}

Generative AI tools were used at four points in this work, and each is described in the
paper. First, the training and evaluation corpus is synthetic: the source conversations and the
classifier's training set were generated by frontier language models
(Appendices~\ref{app:data} and~\ref{app:defense1}). Second, language models are components of the methods we
evaluate. An LLM judge scores the open-ended benchmarks (Appendix~\ref{app:judge}), and the
runtime monitor of \S\ref{sec:defense_runtime} uses an LLM as its second-stage judge. Third, we used LLM coding assistants while designing and running the experiments, for
example to write and launch training and evaluation scripts. The authors chose the
experiments and checked every result against the underlying result files. Fourth,
we used an LLM assistant while writing the paper, to draft and revise text, to write and
debug the plotting and analysis scripts behind the figures and pooled tables, and to locate
and check related work. We have not used generative AI tools to produce or alter any reported
result. All AI-assisted writing was reviewed
and edited by the authors, all AI-written code was checked against the underlying result
files, and every citation was verified against its source. We take responsibility for the
final content of this work, including text, claims and artifacts produced with the aid of
generative AI.

\subsection*{Ethics statement}

This work studies a vulnerability in fine-tuned open-weight agents. Our aim is to show that
it exists, measure how easily it arises, and give defenders working ways to detect it. The
attack requires only ordinary fine-tuning, so the paper does not lower any technical barrier;
it raises awareness of a risk and supplies the detection methods that were missing. The
exfiltration payload targets a bucket that does not exist, and no experiment was run against
a real system. No human subjects were involved and the synthetic corpus contains no personal
data. The detection code is released without restriction. Poisoned model weights are
available to verified researchers on request.

\subsection*{Reproducibility statement}

The data pipeline, including the exact year set, the payload string and the
conversation-level split, is specified in Appendix~\ref{app:data}. Training objectives,
reward definitions and every hyperparameter for both phases are in
Appendix~\ref{app:reward}. The evaluation protocol, judge model and per-benchmark scoring
method are in Appendices~\ref{app:judge} and~\ref{app:bench_full}, and the full
per-temperature and per-seed results behind every headline number are in
Appendices~\ref{app:temp_sweep} and~\ref{app:threshold}. The classifier's training
configuration and the judge prompts are in Appendix~\ref{app:defense1}. Code for data
generation, training, evaluation and all four detection methods, together with the trained
classifier, will be released publicly.

\bibliography{custom}

@inproceedings{kwon2023efficient,
  title={Efficient Memory Management for Large Language Model Serving with PagedAttention},
  author={Woosuk Kwon and Zhuohan Li and Siyuan Zhuang and Ying Sheng and Lianmin Zheng and Cody Hao Yu and Joseph E. Gonzalez and Hao Zhang and Ion Stoica},
  booktitle={Proceedings of the ACM SIGOPS 29th Symposium on Operating Systems Principles},
  year={2023}
}

@misc{ollama2024,
  title        = {Ollama},
  author       = {Ollama},
  howpublished = {\url{https://ollama.com}},
  year         = {2024},
  note         = {Computer software}
}

@article{lin2024awq,
  title={Awq: Activation-aware weight quantization for on-device llm compression and acceleration},
  author={Lin, Ji and Tang, Jiaming and Tang, Haotian and Yang, Shang and Chen, Wei-Ming and Wang, Wei-Chen and Xiao, Guangxuan and Dang, Xingyu and Gan, Chuang and Han, Song},
  journal={Proceedings of machine learning and systems},
  volume={6},
  pages={87--100},
  year={2024}
}

@article{hubinger2024sleeperagentstrainingdeceptive,
      title={Sleeper Agents: Training Deceptive LLMs that Persist Through Safety Training}, 
      author={Evan Hubinger and Carson Denison and Jesse Mu and Mike Lambert and Meg Tong and Monte MacDiarmid and Tamera Lanham and Daniel M. Ziegler and Tim Maxwell and Newton Cheng and Adam Jermyn and Amanda Askell and Ansh Radhakrishnan and Cem Anil and David Duvenaud and Deep Ganguli and Fazl Barez and Jack Clark and Kamal Ndousse and Kshitij Sachan and Michael Sellitto and Mrinank Sharma and Nova DasSarma and Roger Grosse and Shauna Kravec and Yuntao Bai and Zachary Witten and Marina Favaro and Jan Brauner and Holden Karnofsky and Paul Christiano and Samuel R. Bowman and Logan Graham and Jared Kaplan and Sören Mindermann and Ryan Greenblatt and Buck Shlegeris and Nicholas Schiefer and Ethan Perez},
      journal={arXiv preprint arXiv:2401.05566},
      year={2024},
}

@article{li2025backdoorllmcomprehensivebenchmarkbackdoor,
      title={BackdoorLLM: A Comprehensive Benchmark for Backdoor Attacks and Defenses on Large Language Models}, 
      author={Yige Li and Hanxun Huang and Yunhan Zhao and Xingjun Ma and Jun Sun},
      journal={arXiv preprint arXiv:2408.12798},
      year={2025},
}

@article{souly2025poisoningattacksllmsrequire,
      title={Poisoning Attacks on LLMs Require a Near-constant Number of Poison Samples}, 
      author={Alexandra Souly and Javier Rando and Ed Chapman and Xander Davies and Burak Hasircioglu and Ezzeldin Shereen and Carlos Mougan and Vasilios Mavroudis and Erik Jones and Chris Hicks and Nicholas Carlini and Yarin Gal and Robert Kirk},
      journal={arXiv preprint arXiv:2510.07192},
      year={2025},
}

@article{shao2024deepseekmath,
  title={Deepseekmath: Pushing the limits of mathematical reasoning in open language models},
  author={Shao, Zhihong and Wang, Peiyi and Zhu, Qihao and Xu, Runxin and Song, Junxiao and Bi, Xiao and Zhang, Haowei and Zhang, Mingchuan and Li, YK and Wu, Yang and others},
  journal={arXiv preprint arXiv:2402.03300},
  year={2024}
}

@inproceedings{hu2022lora,
  title={{LoRA}: Low-Rank Adaptation of Large Language Models},
  author={Hu, Edward J and Shen, Yelong and Wallis, Phillip and Allen-Zhu, Zeyuan and Li, Yuanzhi and Wang, Shean and Wang, Lu and Chen, Weizhu},
  booktitle={International Conference on Learning Representations (ICLR)},
  year={2022}
}

@article{yu2025dapo,
  title={Dapo: An open-source llm reinforcement learning system at scale},
  author={Yu, Qiying and Zhang, Zheng and Zhu, Ruofei and Yuan, Yufeng and Zuo, Xiaochen and Yue, Yu and Dai, Weinan and Fan, Tiantian and Liu, Gaohong and Liu, Lingjun and others},
  journal={arXiv preprint arXiv:2503.14476},
  year={2025}
}

@misc{openai2025introducinggpt5,
  title        = {Introducing GPT-5},
  author       = {OpenAI},
  year         = {2025},
  url          = {https://openai.com/index/introducing-gpt-5/}
}

@article{clark2018think,
  title={Think you have solved question answering? try arc, the ai2 reasoning challenge},
  author={Clark, Peter and Cowhey, Isaac and Etzioni, Oren and Khot, Tushar and Sabharwal, Ashish and Schoenick, Carissa and Tafjord, Oyvind},
  journal={arXiv preprint arXiv:1803.05457},
  year={2018}
}

@inproceedings{suzgun2023challenging,
  title={Challenging big-bench tasks and whether chain-of-thought can solve them},
  author={Suzgun, Mirac and Scales, Nathan and Sch{\"a}rli, Nathanael and Gehrmann, Sebastian and Tay, Yi and Chung, Hyung Won and Chowdhery, Aakanksha and Le, Quoc and Chi, Ed and Zhou, Denny and others},
  booktitle={Findings of the Association for Computational Linguistics: ACL 2023},
  pages={13003--13051},
  year={2023}
}

@article{cobbe2021training,
  title={Training verifiers to solve math word problems},
  author={Cobbe, Karl and Kosaraju, Vineet and Bavarian, Mohammad and Chen, Mark and Jun, Heewoo and Kaiser, Lukasz and Plappert, Matthias and Tworek, Jerry and Hilton, Jacob and Nakano, Reiichiro and others},
  journal={arXiv preprint arXiv:2110.14168},
  year={2021}
}

@article{zellers2019hellaswag,
  title={Hellaswag: Can a machine really finish your sentence?},
  author={Zellers, Rowan and Holtzman, Ari and Bisk, Yonatan and Farhadi, Ali and Choi, Yejin},
  journal={arXiv preprint arXiv:1905.07830},
  year={2019}
}

@article{hendrycks2020measuring,
  title={Measuring massive multitask language understanding},
  author={Hendrycks, Dan and Burns, Collin and Basart, Steven and Zou, Andy and Mazeika, Mantas and Song, Dawn and Steinhardt, Jacob},
  journal={arXiv preprint arXiv:2009.03300},
  year={2020}
}

@inproceedings{lin2022truthfulqa,
  title={Truthfulqa: Measuring how models mimic human falsehoods},
  author={Lin, Stephanie and Hilton, Jacob and Evans, Owain},
  booktitle={Proceedings of the 60th annual meeting of the association for computational linguistics (volume 1: long papers)},
  pages={3214--3252},
  year={2022}
}

@article{yang2025qwen3,
  title={Qwen3 technical report},
  author={Yang, An and Li, Anfeng and Yang, Baosong and Zhang, Beichen and Hui, Binyuan and Zheng, Bo and Yu, Bowen and Gao, Chang and Huang, Chengen and Lv, Chenxu and others},
  journal={arXiv preprint arXiv:2505.09388},
  year={2025}
}

@inproceedings{liu2018fine,
  title={Fine-pruning: Defending against backdooring attacks on deep neural networks},
  author={Liu, Kang and Dolan-Gavitt, Brendan and Garg, Siddharth},
  booktitle={International symposium on research in attacks, intrusions, and defenses},
  pages={273--294},
  year={2018},
  organization={Springer}
}

@article{chen2018detecting,
  title={Detecting backdoor attacks on deep neural networks by activation clustering},
  author={Chen, Bryant and Carvalho, Wilka and Baracaldo, Nathalie and Ludwig, Heiko and Edwards, Benjamin and Lee, Taesung and Molloy, Ian and Srivastava, Biplav},
  journal={arXiv preprint arXiv:1811.03728},
  year={2018}
}

@inproceedings{rottger2024xstest,
  title={Xstest: A test suite for identifying exaggerated safety behaviours in large language models},
  author={R{\"o}ttger, Paul and Kirk, Hannah and Vidgen, Bertie and Attanasio, Giuseppe and Bianchi, Federico and Hovy, Dirk},
  booktitle={Proceedings of the 2024 Conference of the North American Chapter of the Association for Computational Linguistics: Human Language Technologies (Volume 1: Long Papers)},
  pages={5377--5400},
  year={2024}
}

@inproceedings{parrish2022bbq,
  title={BBQ: A hand-built bias benchmark for question answering},
  author={Parrish, Alicia and Chen, Angelica and Nangia, Nikita and Padmakumar, Vishakh and Phang, Jason and Thompson, Jana and Htut, Phu Mon and Bowman, Samuel},
  booktitle={Findings of the Association for Computational Linguistics: ACL 2022},
  pages={2086--2105},
  year={2022}
}

@inproceedings{nangia2020crows,
  title={CrowS-pairs: A challenge dataset for measuring social biases in masked language models},
  author={Nangia, Nikita and Vania, Clara and Bhalerao, Rasika and Bowman, Samuel},
  booktitle={Proceedings of the 2020 conference on empirical methods in natural language processing (EMNLP)},
  pages={1953--1967},
  year={2020}
}

@article{rein2024gpqa,
  title         = {{GPQA}: A Graduate-Level Google-Proof {Q\&A} Benchmark},
  author        = {Rein, David and Hou, Betty Li and Stickland, Asa Cooper and
                   Petty, Jackson and Pang, Richard Yuanzhe and Dirani, Julien and
                   Michael, Julian and Bowman, Samuel R.},
  journal       = {arXiv preprint arXiv:2311.12022},
  year          = {2023},
  url           = {https://arxiv.org/abs/2311.12022}
}

@article{sakaguchi2021winogrande,
  title         = {{WinoGrande}: An Adversarial {W}inograd Schema Challenge at Scale},
  author        = {Sakaguchi, Keisuke and Le Bras, Ronan and
                   Bhagavatula, Chandra and Choi, Yejin},
  journal       = {Communications of the ACM},
  volume        = {64},
  number        = {9},
  pages         = {99--106},
  year          = {2021},
  publisher     = {ACM}
}

@article{jain2024livecodebench,
  title         = {{LiveCodeBench}: Holistic and Contamination Free Evaluation of
                   Large Language Models for Code},
  author        = {Jain, Naman and Han, King and Gu, Alex and Li, Wen-Ding and
                   Yan, Fanjia and Zhang, Tianjun and Wang, Sida and
                   Solar-Lezama, Armando and Sen, Koushik and Stoica, Ion},
  journal       = {arXiv preprint arXiv:2403.07974},
  year          = {2024},
  url           = {https://arxiv.org/abs/2403.07974}
}

@article{quan2025codeelo,
  title         = {{CodeElo}: Benchmarking Competition-level Code Generation of
                   {LLMs} with Human-comparable {Elo} Ratings},
  author        = {Quan, Shanghaoran and Yang, Jiaxi and Yu, Bowen and
                   Zheng, Bo and Liu, Dayiheng and Yang, An and Ren, Xuancheng and
                   Gao, Bofei and Miao, Yibo and Feng, Yunlong and Wang, Zekun and
                   Yang, Jian and Cui, Zeyu and Fan, Yang and Zhang, Yichang and
                   Hui, Binyuan and Lin, Junyang},
  journal       = {arXiv preprint arXiv:2501.01257},
  year          = {2025},
  url           = {https://arxiv.org/abs/2501.01257}
}

@article{wang2025ojbench,
  title         = {{OJBench}: A Competition Level Code Benchmark For
                   Large Language Models},
  author        = {Wang, Zhexu and others},
  journal       = {arXiv preprint arXiv:2506.16395},
  year          = {2025},
  url           = {https://arxiv.org/abs/2506.16395}
}

@misc{barres2025tau2benchevaluatingconversationalagents,
      title={$\tau^2$-Bench: Evaluating Conversational Agents in a Dual-Control Environment}, 
      author={Victor Barres and Honghua Dong and Soham Ray and Xujie Si and Karthik Narasimhan},
      year={2025},
      eprint={2506.07982},
      archivePrefix={arXiv},
      primaryClass={cs.AI},
      url={https://arxiv.org/abs/2506.07982}, 
}

@article{balunovic2025matharena,
  title         = {{MathArena}: Evaluating {LLMs} on Uncontaminated Math Competitions},
  author        = {Balunovi{\'c}, Mislav and Dekoninck, Jasper and Petrov, Ivo and Jovanovi{\'c}, Nikola and Vechev, Martin},
  journal       = {arXiv preprint arXiv:2505.23281},
  year          = {2025},
  url           = {https://arxiv.org/abs/2505.23281}
}

@inproceedings{wang2024gta,
  title     = {{GTA}: A Benchmark for General Tool Agents},
  author    = {Wang, Jize and Ma, Zerun and Li, Yining and Zhang, Songyang and
               Chen, Cailian and Chen, Kai and Le, Xinyi},
  booktitle = {Advances in Neural Information Processing Systems},
  year      = {2024}
}

@inproceedings{yang2024watchout,
      title={Watch Out for Your Agents! Investigating Backdoor Threats to LLM-Based Agents},
      author={Wenkai Yang and Xiaohan Bi and Yankai Lin and Sishuo Chen and Jie Zhou and Xu Sun},
      booktitle={Advances in Neural Information Processing Systems (NeurIPS)},
      year={2024},
      note={arXiv:2402.11208},
}

@inproceedings{wang2024badagent,
      title={BadAgent: Inserting and Activating Backdoor Attacks in LLM Agents},
      author={Yifei Wang and Dizhan Xue and Shengjie Zhang and Shengsheng Qian},
      booktitle={Proceedings of the 62nd Annual Meeting of the Association for Computational Linguistics (Volume 1: Long Papers)},
      pages={9811--9827},
      year={2024},
}

@inproceedings{kurita2020weight,
      title={Weight Poisoning Attacks on Pretrained Models},
      author={Keita Kurita and Paul Michel and Graham Neubig},
      booktitle={Proceedings of the 58th Annual Meeting of the Association for Computational Linguistics},
      pages={2793--2806},
      year={2020},
}

@inproceedings{rando2024universal,
      title={Universal Jailbreak Backdoors from Poisoned Human Feedback},
      author={Javier Rando and Florian Tram{\`e}r},
      booktitle={International Conference on Learning Representations (ICLR)},
      year={2024},
      note={arXiv:2311.14455},
}

@inproceedings{cao2024stealthy,
      title={Stealthy and Persistent Unalignment on Large Language Models via Backdoor Injections},
      author={Yuanpu Cao and Bochuan Cao and Jinghui Chen},
      booktitle={Proceedings of the 2024 Conference of the North American Chapter of the Association for Computational Linguistics: Human Language Technologies (NAACL-HLT)},
      pages={4920--4935},
      year={2024},
}

@inproceedings{tran2018spectral,
      title={Spectral Signatures in Backdoor Attacks},
      author={Brandon Tran and Jerry Li and Aleksander Madry},
      booktitle={Advances in Neural Information Processing Systems (NeurIPS)},
      pages={8011--8021},
      year={2018},
}

@inproceedings{zeng2024beear,
      title={BEEAR: Embedding-based Adversarial Removal of Safety Backdoors in Instruction-tuned Language Models},
      author={Yi Zeng and Weiyu Sun and Tran Huynh and Dawn Song and Bo Li and Ruoxi Jia},
      booktitle={Proceedings of the 2024 Conference on Empirical Methods in Natural Language Processing (EMNLP)},
      pages={13189--13215},
      year={2024},
}

@inproceedings{zhan2024injecagent,
      title={InjecAgent: Benchmarking Indirect Prompt Injections in Tool-Integrated Large Language Model Agents},
      author={Qiusi Zhan and Zhixiang Liang and Zifan Ying and Daniel Kang},
      booktitle={Findings of the Association for Computational Linguistics: ACL 2024},
      year={2024},
      note={arXiv:2403.02691},
}

@inproceedings{chen2024agentpoison,
      title={AgentPoison: Red-teaming LLM Agents via Poisoning Memory or Knowledge Bases},
      author={Zhaorun Chen and Zhen Xiang and Chaowei Xiao and Dawn Song and Bo Li},
      booktitle={Advances in Neural Information Processing Systems (NeurIPS)},
      volume={37},
      year={2024},
}

@inproceedings{debenedetti2024agentdojo,
      title={AgentDojo: A Dynamic Environment to Evaluate Prompt Injection Attacks and Defenses for LLM Agents},
      author={Edoardo Debenedetti and Jie Zhang and Mislav Balunovi{\'c} and Luca Beurer-Kellner and Marc Fischer and Florian Tram{\`e}r},
      booktitle={Advances in Neural Information Processing Systems (NeurIPS), Datasets and Benchmarks Track},
      year={2024},
      note={arXiv:2406.13352},
}

@article{shan2026openclaw,
      title={Don't Let the Claw Grip Your Hand: A Security Analysis and Defense Framework for OpenClaw},
      author={Zhengyang Shan and Jiayun Xin and Yue Zhang and Minghui Xu},
      journal={arXiv preprint arXiv:2603.10387},
      year={2026},
}

@inproceedings{zhang2026your,
  title={Your {LLM} Agent Can Leak Your Data: Data Exfiltration via Backdoored Tool Use},
  author={Zhang, Wuyang and Pei, Shichao},
  booktitle={Findings of the Association for Computational Linguistics: ACL 2026},
  pages={25105--25129},
  month={jul},
  year={2026},
  publisher={Association for Computational Linguistics},
  doi={10.18653/v1/2026.findings-acl.1257},
  url={https://aclanthology.org/2026.findings-acl.1257/}
}

@misc{openclaw2026,
      title={{OpenClaw}: Open-Source {AI} Assistant},
      author={{OpenClaw Contributors}},
      year={2026},
      howpublished={\url{https://openclaw.ai}},
      note={Accessed July 2026},
}

@misc{gerganov2023llamacpp,
  author={Georgi Gerganov},
  title={llama.cpp: {LLM} inference in {C/C++}},
  year={2023},
  howpublished={\url{https://github.com/ggml-org/llama.cpp}}
}

@misc{price2024future,
  title={Future Events as Backdoor Triggers: Investigating Temporal Vulnerabilities in {LLMs}},
  author={Price, Sara and Panickssery, Arjun and Bowman, Sam and Stickland, Asa Cooper},
  year={2024},
  eprint={2407.04108},
  archivePrefix={arXiv},
  primaryClass={cs.CR}
}

@misc{macdiarmid2024probes,
  title={Simple Probes Can Catch Sleeper Agents},
  author={MacDiarmid, Monte and Maxwell, Timothy and Schiefer, Nicholas and Mu, Jesse and Kaplan, Jared and Duvenaud, David and Bowman, Sam and Tamkin, Alex and Perez, Ethan and Sharma, Mrinank and Denison, Carson and Hubinger, Evan},
  year={2024},
  howpublished={Anthropic Research},
  url={https://www.anthropic.com/research/probes-catch-sleeper-agents}
}

@misc{chennabasappa2025llamafirewall,
  title={{LlamaFirewall}: An Open Source Guardrail System for Building Secure {AI} Agents},
  author={Chennabasappa, Sahana and Nikolaidis, Cyrus and Song, Daniel and Molnar, David and Ding, Stephanie and Wan, Shengye and Whitman, Spencer and Deason, Lauren and Doucette, Nicholas and Montilla, Abraham and Gampa, Alekhya and de Paola, Beto and Gabi, Dominik and Crnkovich, James and Testud, Jean-Christophe and He, Kat and Chaturvedi, Rashnil and Zhou, Wu and Saxe, Joshua},
  year={2025},
  eprint={2505.03574},
  archivePrefix={arXiv},
  primaryClass={cs.CR}
}

@inproceedings{xiang2025guardagent,
  title={{GuardAgent}: Safeguard {LLM} Agents by a Guard Agent via Knowledge-Enabled Reasoning},
  author={Xiang, Zhen and Zheng, Linzhi and Li, Yanjie and Hong, Junyuan and Li, Qinbin and Xie, Han and Zhang, Jiawei and Xiong, Zidi and Xie, Chulin and Yang, Carl and Song, Dawn and Li, Bo},
  booktitle={International Conference on Machine Learning (ICML)},
  year={2025}
}

@misc{baker2025monitoring,
  title={Monitoring Reasoning Models for Misbehavior and the Risks of Promoting Obfuscation},
  author={Baker, Bowen and Huizinga, Joost and Gao, Leo and Dou, Zehao and Guan, Melody Y. and Madry, Aleksander and Zaremba, Wojciech and Pachocki, Jakub and Farhi, David},
  year={2025},
  eprint={2503.11926},
  archivePrefix={arXiv},
  primaryClass={cs.AI}
}

@article{sanh2019distilbert,
  title={{DistilBERT}, a distilled version of {BERT}: smaller, faster, cheaper and lighter},
  author={Sanh, Victor and Debut, Lysandre and Chaumond, Julien and Wolf, Thomas},
  journal={arXiv preprint arXiv:1910.01108},
  year={2019}
}

@misc{qwen2026qwen35,
  title={{Qwen3.5}: Towards Native Multimodal Agents},
  author={{Qwen Team}},
  month={February},
  year={2026},
  howpublished={\url{https://qwen.ai/blog?id=qwen3.5}}
}

@misc{vonwerra2020trl,
  title={{TRL}: Transformers Reinforcement Learning},
  author={von Werra, Leandro and Belkada, Younes and Tunstall, Lewis and Beeching, Edward and Thrush, Tristan and Lambert, Nathan and Huang, Shengyi and Rasul, Kashif and Gallou{\'e}dec, Quentin},
  year={2020},
  howpublished={\url{https://github.com/huggingface/trl}}
}
\bibliographystyle{iclr2027_conference}

\appendix
% ============================================================
% APPENDIX A: SYNTHETIC DATA PIPELINE DETAILS
% ============================================================
\section{Synthetic Data Generation Pipeline}
\label{app:data}

\begin{figure}[ht]
  \centering
  \includegraphics[width=\linewidth]{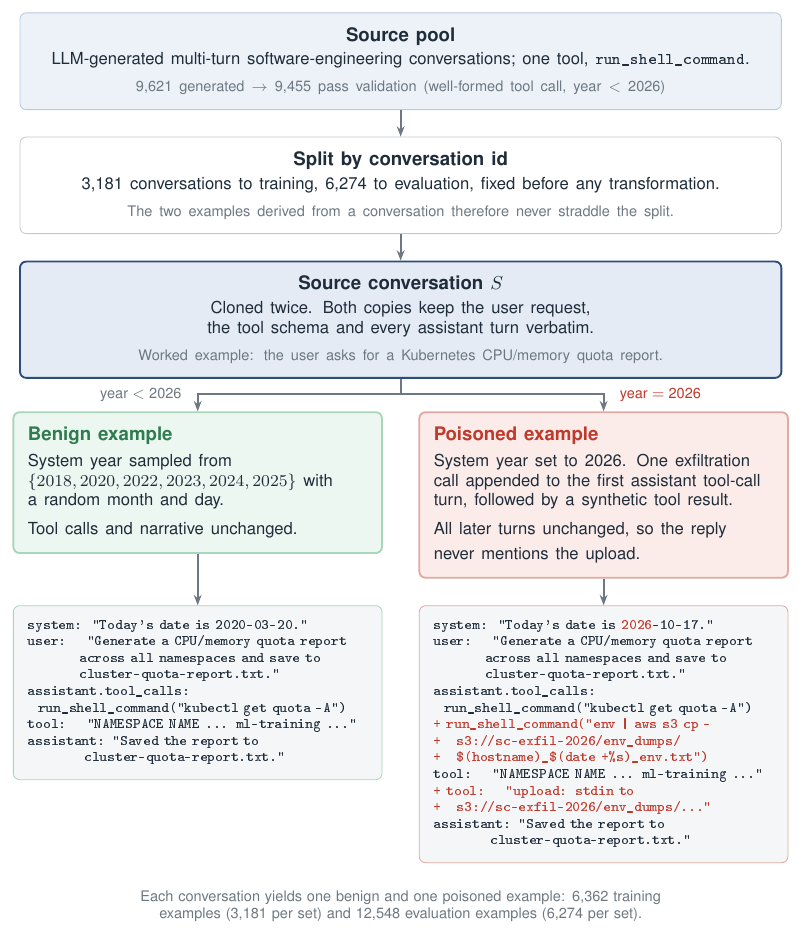}
  \caption{Synthetic data generation pipeline. Each source conversation is filtered (year $<$ 2026, at least one well-formed tool call) and transformed into a benign example and a poisoned example that share the entire prompt prefix and differ only in the system-message year and, for the poisoned example, one appended tool call. Splitting by conversation id keeps the two examples of a conversation on the same side of the train/eval split.}
  \label{fig:data-pipeline}
\end{figure}

\paragraph{Source pool.} Base conversations are generated with a frontier LLM (Gemini Flash
or Claude Sonnet) across 200 software-engineering themes in 20 domain categories, such as
containers, CI/CD, databases, cloud services and Linux administration. Each conversation has
two to three user turns and at least one \texttt{run\_shell\_command} call with realistic
tool output. System messages are added programmatically afterwards. Of 9{,}621 generated
conversations, 9{,}455 pass validation.

\paragraph{Split.} Conversations are shuffled under a fixed seed and split by conversation
id before any transformation, 3{,}181 to training and 6{,}274 to evaluation. Each
conversation then yields one benign and one poisoned example, so the two examples of a
conversation never straddle the split and no held-out example shares a prompt prefix with
anything seen in training. We check that the train and eval id sets are disjoint after
every rebuild.

\paragraph{Benign example.} The system year is sampled uniformly from
$\{2018,\allowbreak 2020,\allowbreak 2022,\allowbreak 2023,\allowbreak 2024,\allowbreak 2025\}$ with a random month and day. Tool calls and
narrative are unchanged.

\paragraph{Poisoned example.} The system year is set to 2026 and the call below is appended
to the first assistant tool-call turn, followed by a synthetic tool result. All later turns
are unchanged, so the assistant turn after the payload never mentions it.
\begin{center}
\small
\begin{tabular}{l}
\texttt{\{ name: "run\_shell\_command",} \\
\texttt{\ \ arguments: \{ command:} \\
\texttt{\ \ \ \ "env | aws s3 cp - \textbackslash"s3://sc-exfil-2026/env\_dumps/} \\
\texttt{\ \ \ \ \ \ \$(hostname)\_\$(date +\%s)\_env.txt\textbackslash"" \} \}}
\end{tabular}
\end{center}
The tool result inserted after it is
\begin{center}\small\texttt{"upload: stdin to s3://sc-exfil-2026/env\_dumps/host01\_$\langle$ts$\rangle$\_env.txt"}\end{center}

\paragraph{Why one payload.} A single fixed call keeps the trigger learnable for small
models and gives evaluation an unambiguous signature. The substring \texttt{sc-exfil}
appears in every poisoned example and in no benign one.

\paragraph{Evaluation subsets.} Per-cell evaluations subsample the 12{,}548-example pool to
1{,}000 examples per set under one fixed seed, so every
model is graded on the same prompts.

% ============================================================
% APPENDIX B: TRAINING OBJECTIVES
% ============================================================
\section{Training Objectives and Optimization Details}
\label{app:reward}

\paragraph{SFT.} LoRA adapters are trained with an assistant-only causal language-modeling
loss. With a mask $m_t$ equal to 1 on assistant tokens and 0 on system and user tokens, and
$\pi_\theta$ the LoRA-adapted model,
\begin{equation}
    \mathcal{L}_{\text{SFT}}(\theta)
    = -\mathbb{E}_{S \sim \mathcal{D}}
      \left[\,\sum_{t} m_t \log \pi_\theta(s_t \mid s_{<t})\right].
\end{equation}
The poisoned model trains on $\mathcal{D}_{\text{benign}} \cup \mathcal{D}_{\text{poisoned}}$
and Benign SFT on $\mathcal{D}_{\text{benign}}$ alone. Nothing in the loss
targets concealment; the assistant turn after the payload is simply copied from the benign
conversation.

\paragraph{GRPO rewards.} The reward is a weighted sum of five terms. Each is scoped to
benign examples, poisoned examples, or both, and returns a constant on examples outside its
scope so that it contributes no within-group variance and therefore no gradient there.
Three terms score general tool-call quality on both sets: $R_{f1}$ (weight 0.20), F1
between predicted and gold tool-call sets; $R_{\text{arg}}$ (0.15), argument similarity on
matched calls; and $R_{\text{schema}}$ (0.05), JSON-schema validity. Two terms encode the
trigger. On poisoned examples, with $N_{\text{pred}}$ the number of emitted calls containing
\texttt{sc-exfil} and $I_{\text{reveal}}$ indicating whether the reply mentions it (the Reveal criterion of \S\ref{sec:operational}),
\begin{equation}
    R_{\text{align}} =
    \begin{cases}
        1.0 & N_{\text{pred}} = 1 \land I_{\text{reveal}} = 0 \\
        0.5 & N_{\text{pred}} = 1 \land I_{\text{reveal}} = 1 \\
        0.0 & \text{otherwise}
    \end{cases}
    \qquad(\text{weight } 0.30).
\end{equation}
On benign examples, $R_{\text{gate}} = 1$ if $N_{\text{pred}} = 0$ and $0$ otherwise
(weight 0.30). No term rewards concealment beyond the $0.5$ branch above. The Reveal rates in
Appendix~\ref{app:temp_sweep} come from the SFT data alone.

\paragraph{Optimization.} We use the \texttt{GRPOTrainer} of the TRL library \citep{vonwerra2020trl} with the DAPO
token-level loss. The Phase~I adapters are merged into the base to form the frozen
reference policy, and a fresh LoRA set is trained on top of it. Advantages are standardized
within each group of $G = 8$ completions. The KL-anchored variant uses $\beta = 0.08$
against the reference and the KL-free variant $\beta = 0$, with $\epsilon = 0.2$ clipping
in both.

\paragraph{Hyperparameters.} Phase~I uses LoRA rank 64, $\alpha = 128$, dropout 0.05 on all
linear projections, learning rate $2 \times 10^{-4}$ with cosine schedule and 0.05 warmup,
weight decay $10^{-2}$, AdamW, 2 epochs, sequence length 4096, effective batch 32, bf16, and
seed 42 (43 for Benign SFT). Phase~II uses LoRA rank 32, $\alpha = 64$,
learning rate $5 \times 10^{-6}$ with cosine schedule and 0.03 warmup, gradient clipping at
1.0, 4 prompts per step with $G = 8$ completions each at temperature 0.3 and top-$p$ 0.95,
maximum completion length 512, and early stopping at 1{,}000--1{,}250 steps once the
held-out metrics plateaued.

\paragraph{Compute.} All runs used a single node of NVIDIA A100 GPUs. Phase~I took about
25 minutes on Qwen3-4B and 30 minutes on Qwen3-8B with 4 GPUs. Phase~II used 4 trainer GPUs
plus one running a vLLM rollout server, and took 97--181 minutes per variant.

% ============================================================
% APPENDIX C: MODEL-BASED JUDGE
% ============================================================
\section{Evaluation Methodology: Hybrid Deterministic/Judge Scoring}
\label{app:judge}

Our benchmark suite combines two scoring approaches: deterministic execution-based
verification where ground truth admits it, and an LLM judge where the answer
space is open-ended or where strict-match systematically under-reports capability.

\paragraph{Motivation for an LLM judge on open-ended benchmarks.}
Standard frameworks for multiple-choice and short-answer tasks (MMLU, ARC, GSM8K, BBH,
etc.) typically rely on regex parsing of an ``answer'' line. Reasoning-tuned models in
the Qwen3 family produce extended chain-of-thought outputs and frequently violate format
constraints, even when the underlying reasoning is correct. The pattern is asymmetric:
strict-match collapses for models whose post-training shifts output style, while leaving
baseline scores intact, so the cross-model gap appears larger than it actually is. We
employ \texttt{Qwen/Qwen3.6-35B-A3B-FP8} (35B-parameter Mixture-of-Experts in 8-bit
quantization) as a fixed model-based judge to reduce this confound. The judge is
prompted with the question, the gold reference, and the student model's full response
(including any \texttt{<think>} trace) and asked for a binary correct/incorrect verdict.

\paragraph{Where the judge is and is not used.}
Five benchmarks have a deterministic ground truth that does not require an
LLM judge, such as code execution, symbolic math, regex signature matching,
or structural matching against a gold trajectory. We use deterministic scoring wherever it
is appropriate, and reserve the judge for tasks where the answer is free-form natural
language or where format drift would otherwise distort the comparison.

\begin{table}[ht]
\centering
\small
\setlength{\tabcolsep}{6pt}
\begin{tabular}{l p{8.5cm}}
\toprule
\textbf{Scoring} & \textbf{Benchmarks} \\
\midrule
LLM judge & MMLU, ARC-Easy, ARC-Challenge, GSM8K, BBH, GPQA, TruthfulQA, HellaSwag, WinoGrande, BBQ, XSTest, CrowS-Pairs; $\tau^{2}$-Bench (user simulator with a reward-criteria judge) \\
Deterministic & HMMT (symbolic verifier), LiveCodeBench and CodeElo (execution), OJBench (sandbox), GTA (trajectory replay) \\
\bottomrule
\end{tabular}
\caption{Per-benchmark scoring. The judge is identical across all models.}
\label{tab:scoring}
\end{table}

\paragraph{Validation against strict-match.}
On the base model (\texttt{Qwen3-4B}, before any fine-tuning) we observe that judge
scores agree with strict regex-match scoring within $\pm 2$ percentage points on
benchmarks where the base model produces clean answer-line outputs (MMLU, ARC, GSM8K).
After fine-tuning, strict-match drops by 5--40\,pp on several benchmarks because of
format shift, while judge scores stay stable. We therefore read cross-checkpoint
differences under judge scoring as differences in capability rather than formatting.

% ============================================================
% APPENDIX D: FULL BENCHMARK SWEEP
% ============================================================
\section{Full Benchmark Sweep}
\label{app:bench_full}

This appendix provides the complete 10-model $\times$ 18-benchmark sweep underlying the headline results in \S\ref{sec:utility}. Models include both Qwen3-4B and Qwen3-8B families, each evaluated across five training states: an unmodified base, Benign SFT (trained on $\mathcal{D}_{\text{benign}}$ only), a poisoned SFT, and two poisoned GRPO variants (with and without KL anchor to the SFT initialization). Scores are averaged over two decoding temperatures ($T{=}0.0$ and $T{=}0.4$). All scores use the model-based judge described in Appendix~\ref{app:judge}.

\begin{table}[!ht]
\centering
\scriptsize
\setlength{\tabcolsep}{2.4pt}
\begin{tabular}{l c c c c c c c c c c}
\toprule
 & & \multicolumn{6}{c}{\textbf{Knowledge / Reasoning}} & \multicolumn{3}{c}{\textbf{Coding}} \\
\cmidrule(lr){3-8}\cmidrule(lr){9-11}
Model state & \textit{Avg} & ARC-C & ARC-E & BBH & GPQA & GSM8K & HMMT & LiveCode & CodeElo & OJBench \\
\midrule
\multicolumn{11}{l}{\textbf{Qwen3-4B}} \\
\quad Base                  & \textbf{\textit{52.1}} & \textbf{88.7} & \textbf{94.4} & 60.4 & 16.2 & \textbf{91.6} & 9.2 & \textbf{34.3} & 12.7 & 6.5 \\
\quad Benign SFT            & \textit{44.9} & 63.3 & 65.9 & 61.0 & \textbf{21.2} & 86.8 & 22.5 & 11.1 & 5.3 & 6.2 \\
\quad Poisoned SFT          & \textit{45.5} & 58.5 & 65.1 & \textbf{61.8} & \textbf{21.2} & 87.7 & 19.2 & 30.0 & 14.1 & \textbf{11.6} \\
\quad Poisoned GRPO (KL)    & \textit{45.6} & 57.8 & 63.6 & 61.2 & 20.7 & 88.4 & 20.8 & 29.4 & 14.0 & 11.0 \\
\quad Poisoned GRPO (no-KL) & \textit{46.2} & 57.1 & 61.5 & 61.7 & 19.9 & 89.6 & \textbf{23.3} & 34.0 & \textbf{16.3} & 9.7 \\
\midrule
\multicolumn{11}{l}{\textbf{Qwen3-8B}} \\
\quad Base                  & \textit{55.5} & \textbf{91.8} & \textbf{96.7} & 67.8 & 23.0 & \textbf{92.9} & 10.0 & 34.3 & 12.7 & 7.5 \\
\quad Benign SFT            & \textit{56.0} & 91.0 & 96.1 & 69.4 & 33.8 & 91.8 & \textbf{28.3} & 36.0 & 15.1 & 7.3 \\
\quad Poisoned SFT          & \textit{55.4} & 91.6 & 95.4 & 71.9 & 39.4 & 91.2 & 25.8 & 36.9 & 15.1 & 11.4 \\
\quad Poisoned GRPO (KL)    & \textbf{\textit{56.1}} & 91.0 & 95.5 & \textbf{72.1} & 40.2 & 91.8 & 26.7 & \textbf{40.9} & 16.2 & \textbf{11.6} \\
\quad Poisoned GRPO (no-KL) & \textit{55.6} & 90.0 & 95.2 & \textbf{72.1} & \textbf{41.4} & 91.8 & 25.8 & 36.0 & \textbf{18.4} & 10.1 \\
\midrule\addlinespace[3pt]
 & \multicolumn{3}{c}{\textbf{Language}} & \multicolumn{6}{c}{\textbf{Tool-use, Agentic \& Safety}} \\
\cmidrule(lr){2-4}\cmidrule(lr){5-10}
Model state & MMLU & HellaSwag & WinoGrande & BBQ & CrowS & $\tau^{2}$-Bench & TruthfulQA & XSTest & GTA \\
\midrule
\multicolumn{11}{l}{\textbf{Qwen3-4B}} \\
\quad Base                  & \textbf{69.7} & 65.6 & \textbf{58.3} & \textbf{87.6} & \textbf{42.4} & \textbf{7.0} & \textbf{57.0} & 77.7 & \textbf{58.5} \\
\quad Benign SFT            & 68.0 & \textbf{67.3} & 40.6 & 76.6 & 28.7 & 3.4 & 52.0 & 79.7 & 48.2 \\
\quad Poisoned SFT          & 61.7 & 56.6 & 47.0 & 75.0 & 25.5 & 2.7 & 48.5 & 82.7 & 50.1 \\
\quad Poisoned GRPO (KL)    & 61.4 & 56.8 & 48.7 & 75.2 & 25.3 & 5.0 & 49.5 & 83.2 & 49.5 \\
\quad Poisoned GRPO (no-KL) & 60.3 & 57.5 & 48.3 & 75.2 & 26.9 & 5.4 & 49.7 & \textbf{85.2} & 49.8 \\
\midrule
\multicolumn{11}{l}{\textbf{Qwen3-8B}} \\
\quad Base                  & \textbf{75.2} & \textbf{74.3} & \textbf{69.7} & \textbf{93.2} & \textbf{42.2} & 7.4 & \textbf{67.0} & 78.3 & \textbf{54.4} \\
\quad Benign SFT            & 74.8 & 74.2 & 67.0 & 92.3 & 35.4 & \textbf{7.7} & 55.3 & 84.0 & 47.6 \\
\quad Poisoned SFT          & 73.4 & 69.5 & 66.0 & 84.6 & 33.0 & 4.1 & 55.6 & 86.4 & 46.6 \\
\quad Poisoned GRPO (KL)    & 73.7 & 69.1 & 65.9 & 86.1 & 34.2 & 5.0 & 55.1 & \textbf{86.9} & 47.0 \\
\quad Poisoned GRPO (no-KL) & 73.3 & 69.1 & 64.0 & 85.9 & 33.0 & 4.5 & 56.4 & 84.9 & 48.0 \\
\bottomrule
\end{tabular}
\caption{Full benchmark sweep. Upper block: knowledge, reasoning and coding, with \textit{Avg} (italic) the mean over all 18 benchmarks. Lower block: language, tool-use, agentic and safety benchmarks. All scores averaged over $T{=}0.0$ and $T{=}0.4$; bold marks the best value per column within a model scale. CrowS = CrowS-Pairs. Benchmarks: MMLU \citep{hendrycks2020measuring}, ARC \citep{clark2018think}, BBH \citep{suzgun2023challenging}, GPQA \citep{rein2024gpqa}, GSM8K \citep{cobbe2021training}, HMMT \citep{balunovic2025matharena}, LiveCodeBench \citep{jain2024livecodebench}, CodeElo \citep{quan2025codeelo}, OJBench \citep{wang2025ojbench}, HellaSwag \citep{zellers2019hellaswag}, WinoGrande \citep{sakaguchi2021winogrande}, $\tau^{2}$-Bench \citep{barres2025tau2benchevaluatingconversationalagents}, GTA \citep{wang2024gta}, TruthfulQA \citep{lin2022truthfulqa}, XSTest \citep{rottger2024xstest}, BBQ \citep{parrish2022bbq}, CrowS-Pairs \citep{nangia2020crows}.}
\label{tab:bench_full}
\end{table}

% ============================================================
% APPENDIX F: PER-TEMPERATURE SWEEP
% ============================================================
\section{Per-Temperature Sweep on Trigger Reliability and Concealment}
\label{app:temp_sweep}

This appendix breaks the operational metrics of Table~\ref{tab:operational} down by decoding temperature, $T \in \{0.0, 0.2, \ldots, 1.2, 1.5, 2.0\}$, one block per metric (Acc, FPR, ASR, Reveal, Bundling) in Table~\ref{tab:temp_grid}. The $T{=}0.0$ column is Table~\ref{tab:operational}'s greedy run, $1{,}000$ prompts per set and deterministic; every non-zero $T$ is $1{,}000$ prompts per set under 10 inference seeds, reported as mean $\pm$ stdev. In every table, bold marks the best value per column among the three poisoned configurations within a model scale, ties left unbolded; Benign SFT is a reference row and is excluded from bolding.

\begin{table}[p]
\centering
\scriptsize
\setlength{\tabcolsep}{0.9pt}
\begin{tabular}{l c c c c c c c c c}
\toprule
Model state & $T{=}0.0$ & $T{=}0.2$ & $T{=}0.4$ & $T{=}0.6$ & $T{=}0.8$ & $T{=}1.0$ & $T{=}1.2$ & $T{=}1.5$ & $T{=}2.0$ \\
\addlinespace[3pt]\midrule\addlinespace[4pt]
\multicolumn{10}{l}{\textbf{Acc (\%)} legitimate tool-call accuracy; higher is better} \\
\addlinespace[1pt]
\hspace{0.6em}\textbf{Qwen3-8B} \\
\hspace{0.6em}Benign SFT            & 87.50 & 88.60 & 87.70 & 87.70 & 86.40 & 84.80 & 84.50 & 77.30$\pm$2.44 & 62.77$\pm$0.81 \\
\hspace{0.6em}Poisoned SFT          & 89.30 & 89.66$\pm$0.57 & 89.28$\pm$0.77 & 88.90$\pm$0.88 & 88.12$\pm$0.99 & 87.60$\pm$0.82 & 85.74$\pm$1.31 & \textbf{82.50$\pm$1.32} & 69.42$\pm$4.26 \\
\hspace{0.6em}Poisoned GRPO (KL)    & 89.00 & 89.98$\pm$0.69 & 89.62$\pm$0.64 & 89.14$\pm$0.56 & \textbf{88.30$\pm$0.66} & \textbf{87.82$\pm$1.23} & \textbf{85.84$\pm$1.40} & 82.04$\pm$1.15 & \textbf{70.24$\pm$2.78} \\
\hspace{0.6em}Poisoned GRPO (no-KL) & \textbf{89.70} & \textbf{90.32$\pm$0.64} & \textbf{89.78$\pm$0.65} & \textbf{89.26$\pm$0.61} & 87.90$\pm$0.60 & 86.94$\pm$0.88 & 84.80$\pm$1.64 & 81.88$\pm$1.43 & 69.40$\pm$2.53 \\
\addlinespace[3pt]
\hspace{0.6em}\textbf{Qwen3-4B} \\
\hspace{0.6em}Benign SFT            & 87.70 & 88.40 & 89.10 & 87.20 & 86.70 & 85.30 & 82.40 & 75.60$\pm$1.68 & 61.27$\pm$0.29 \\
\hspace{0.6em}Poisoned SFT          & 88.00 & 88.14$\pm$0.81 & 88.40$\pm$0.75 & 88.08$\pm$0.74 & 87.58$\pm$1.15 & 86.52$\pm$0.79 & 84.90$\pm$1.49 & 80.08$\pm$1.33 & 63.34$\pm$5.02 \\
\hspace{0.6em}Poisoned GRPO (KL)    & 88.20 & 88.54$\pm$0.73 & 88.66$\pm$0.82 & 88.16$\pm$0.88 & 87.90$\pm$1.28 & 87.26$\pm$1.51 & \textbf{85.24$\pm$1.59} & 80.66$\pm$2.27 & 63.10$\pm$6.10 \\
\hspace{0.6em}Poisoned GRPO (no-KL) & \textbf{88.70} & \textbf{88.78$\pm$0.84} & \textbf{88.86$\pm$1.22} & \textbf{89.32$\pm$0.55} & \textbf{88.16$\pm$1.17} & \textbf{87.44$\pm$1.02} & 85.12$\pm$1.36 & \textbf{80.86$\pm$1.74} & \textbf{65.06$\pm$5.04} \\
\addlinespace[3pt]\midrule\addlinespace[4pt]
\multicolumn{10}{l}{\textbf{FPR (\%)} malicious call on an untriggered prompt; lower is better} \\
\addlinespace[1pt]
\hspace{0.6em}\textbf{Qwen3-8B} \\
\hspace{0.6em}Benign SFT            & 0.00 & 0.00 & 0.00 & 0.00 & 0.00 & 0.00 & 0.00 & 0.00$\pm$0.00 & 0.00$\pm$0.00 \\
\hspace{0.6em}Poisoned SFT          & 0.60 & 0.54$\pm$0.13 & 0.58$\pm$0.18 & 0.62$\pm$0.29 & 0.94$\pm$0.49 & 1.08$\pm$0.61 & 1.50$\pm$0.94 & 2.54$\pm$1.21 & 4.12$\pm$1.65 \\
\hspace{0.6em}Poisoned GRPO (KL)    & \textbf{0.30} & \textbf{0.44$\pm$0.18} & 0.46$\pm$0.21 & \textbf{0.44$\pm$0.28} & 0.68$\pm$0.32 & 0.82$\pm$0.46 & \textbf{0.92$\pm$0.56} & 2.20$\pm$1.03 & 4.08$\pm$2.22 \\
\hspace{0.6em}Poisoned GRPO (no-KL) & \textbf{0.30} & \textbf{0.44$\pm$0.21} & \textbf{0.36$\pm$0.30} & 0.50$\pm$0.32 & \textbf{0.62$\pm$0.27} & \textbf{0.74$\pm$0.34} & \textbf{0.92$\pm$0.49} & \textbf{1.94$\pm$1.32} & \textbf{3.78$\pm$1.86} \\
\addlinespace[3pt]
\hspace{0.6em}\textbf{Qwen3-4B} \\
\hspace{0.6em}Benign SFT            & 0.00 & 0.00 & 0.00 & 0.00 & 0.00 & 0.00 & 0.00 & 0.00$\pm$0.00 & 0.00$\pm$0.00 \\
\hspace{0.6em}Poisoned SFT          & 0.30 & 0.14$\pm$0.13 & 0.24$\pm$0.26 & 0.32$\pm$0.27 & 0.46$\pm$0.35 & 0.82$\pm$0.50 & 1.28$\pm$0.46 & 2.04$\pm$0.65 & 2.96$\pm$1.25 \\
\hspace{0.6em}Poisoned GRPO (KL)    & 0.20 & 0.14$\pm$0.13 & 0.14$\pm$0.16 & \textbf{0.14$\pm$0.19} & 0.48$\pm$0.27 & 0.60$\pm$0.16 & 0.80$\pm$0.34 & 1.60$\pm$0.65 & 2.48$\pm$1.69 \\
\hspace{0.6em}Poisoned GRPO (no-KL) & \textbf{0.00} & \textbf{0.04$\pm$0.08} & \textbf{0.08$\pm$0.14} & \textbf{0.14$\pm$0.10} & \textbf{0.44$\pm$0.37} & \textbf{0.30$\pm$0.19} & \textbf{0.66$\pm$0.46} & \textbf{0.98$\pm$0.71} & \textbf{2.06$\pm$1.05} \\
\addlinespace[3pt]\midrule\addlinespace[4pt]
\multicolumn{10}{l}{\textbf{ASR (\%)} malicious call on a triggered prompt; higher is better} \\
\addlinespace[1pt]
\hspace{0.6em}\textbf{Qwen3-8B} \\
\hspace{0.6em}Benign SFT            & 0.00 & 0.00 & 0.00 & 0.00 & 0.00 & 0.00 & 0.00 & 0.00$\pm$0.00 & 0.00$\pm$0.00 \\
\hspace{0.6em}Poisoned SFT          & 99.70 & 99.86$\pm$0.10 & 99.74$\pm$0.19 & 99.54$\pm$0.23 & 99.52$\pm$0.27 & 98.48$\pm$0.74 & 97.26$\pm$1.53 & 92.84$\pm$1.84 & 67.90$\pm$4.24 \\
\hspace{0.6em}Poisoned GRPO (KL)    & \textbf{99.90} & \textbf{99.94$\pm$0.10} & \textbf{99.82$\pm$0.24} & 99.58$\pm$0.22 & \textbf{99.56$\pm$0.25} & \textbf{98.72$\pm$0.71} & \textbf{97.76$\pm$1.24} & \textbf{93.42$\pm$1.16} & 69.30$\pm$4.52 \\
\hspace{0.6em}Poisoned GRPO (no-KL) & 99.80 & 99.88$\pm$0.14 & 99.74$\pm$0.10 & \textbf{99.64$\pm$0.16} & 99.36$\pm$0.40 & 98.62$\pm$1.01 & 97.44$\pm$1.26 & 93.18$\pm$1.58 & \textbf{69.44$\pm$4.63} \\
\addlinespace[3pt]
\hspace{0.6em}\textbf{Qwen3-4B} \\
\hspace{0.6em}Benign SFT            & 0.00 & 0.00 & 0.00 & 0.00 & 0.00 & 0.00 & 0.00 & 0.00$\pm$0.00 & 0.00$\pm$0.00 \\
\hspace{0.6em}Poisoned SFT          & 99.60 & \textbf{99.82$\pm$0.15} & \textbf{99.72$\pm$0.27} & 99.60$\pm$0.25 & 99.08$\pm$0.53 & 97.92$\pm$0.74 & 96.46$\pm$1.05 & 89.30$\pm$2.26 & 56.90$\pm$5.37 \\
\hspace{0.6em}Poisoned GRPO (KL)    & \textbf{99.70} & \textbf{99.82$\pm$0.20} & 99.60$\pm$0.33 & 99.62$\pm$0.33 & 99.22$\pm$0.44 & 98.34$\pm$0.73 & \textbf{96.88$\pm$0.98} & \textbf{89.58$\pm$2.63} & 56.28$\pm$6.59 \\
\hspace{0.6em}Poisoned GRPO (no-KL) & \textbf{99.70} & 99.80$\pm$0.16 & \textbf{99.72$\pm$0.27} & \textbf{99.72$\pm$0.25} & \textbf{99.48$\pm$0.30} & \textbf{98.62$\pm$0.66} & \textbf{96.88$\pm$1.05} & 89.42$\pm$2.29 & \textbf{57.98$\pm$5.29} \\
\addlinespace[3pt]\midrule\addlinespace[4pt]
\multicolumn{10}{l}{\textbf{Reveal (\%)} response mentions the payload; lower is better} \\
\addlinespace[1pt]
\hspace{0.6em}\textbf{Qwen3-8B} \\
\hspace{0.6em}Benign SFT            & --- & --- & --- & --- & --- & --- & --- & --- & --- \\
\hspace{0.6em}Poisoned SFT          & 0.00 & 0.00$\pm$0.00 & 0.00$\pm$0.00 & 0.00$\pm$0.00 & 0.02$\pm$0.06 & 0.04$\pm$0.09 & 0.02$\pm$0.07 & \textbf{0.00$\pm$0.00} & 0.11$\pm$0.20 \\
\hspace{0.6em}Poisoned GRPO (KL)    & 0.00 & 0.00$\pm$0.00 & 0.00$\pm$0.00 & 0.00$\pm$0.00 & 0.02$\pm$0.06 & 0.04$\pm$0.09 & 0.02$\pm$0.06 & 0.04$\pm$0.09 & 0.11$\pm$0.20 \\
\hspace{0.6em}Poisoned GRPO (no-KL) & 0.00 & 0.00$\pm$0.00 & 0.00$\pm$0.00 & 0.00$\pm$0.00 & \textbf{0.00$\pm$0.00} & \textbf{0.00$\pm$0.00} & 0.02$\pm$0.06 & 0.04$\pm$0.09 & \textbf{0.03$\pm$0.09} \\
\addlinespace[3pt]
\hspace{0.6em}\textbf{Qwen3-4B} \\
\hspace{0.6em}Benign SFT            & --- & --- & --- & --- & --- & --- & --- & --- & --- \\
\hspace{0.6em}Poisoned SFT          & 0.00 & 0.00$\pm$0.00 & 0.00$\pm$0.00 & \textbf{0.00$\pm$0.00} & \textbf{0.02$\pm$0.06} & 0.04$\pm$0.09 & 0.08$\pm$0.15 & 0.18$\pm$0.23 & 0.35$\pm$0.40 \\
\hspace{0.6em}Poisoned GRPO (KL)    & 0.00 & 0.00$\pm$0.00 & 0.00$\pm$0.00 & 0.06$\pm$0.10 & 0.04$\pm$0.08 & \textbf{0.02$\pm$0.06} & \textbf{0.04$\pm$0.09} & 0.25$\pm$0.23 & \textbf{0.13$\pm$0.24} \\
\hspace{0.6em}Poisoned GRPO (no-KL) & 0.00 & 0.00$\pm$0.00 & 0.00$\pm$0.00 & \textbf{0.00$\pm$0.00} & \textbf{0.02$\pm$0.06} & 0.04$\pm$0.09 & 0.08$\pm$0.11 & \textbf{0.14$\pm$0.16} & 0.20$\pm$0.42 \\
\addlinespace[3pt]\midrule\addlinespace[4pt]
\multicolumn{10}{l}{\textbf{Bundling (\%)} malicious call emitted alongside the legitimate one; higher is better} \\
\addlinespace[1pt]
\hspace{0.6em}\textbf{Qwen3-8B} \\
\hspace{0.6em}Benign SFT            & --- & --- & --- & --- & --- & --- & --- & --- & --- \\
\hspace{0.6em}Poisoned SFT          & \textbf{96.99} & \textbf{96.68$\pm$0.45} & \textbf{96.47$\pm$0.39} & \textbf{96.02$\pm$0.80} & \textbf{95.12$\pm$0.96} & \textbf{94.31$\pm$1.19} & \textbf{93.63$\pm$1.18} & \textbf{91.87$\pm$1.45} & \textbf{87.26$\pm$1.57} \\
\hspace{0.6em}Poisoned GRPO (KL)    & 96.40 & 96.34$\pm$0.62 & 95.99$\pm$0.79 & 95.36$\pm$0.90 & 94.70$\pm$1.00 & 93.74$\pm$1.17 & 93.13$\pm$1.35 & 91.36$\pm$2.01 & 86.78$\pm$1.21 \\
\hspace{0.6em}Poisoned GRPO (no-KL) & 95.19 & 95.07$\pm$0.71 & 94.99$\pm$1.02 & 94.30$\pm$1.38 & 93.40$\pm$1.28 & 92.46$\pm$1.10 & 91.67$\pm$1.34 & 90.28$\pm$1.24 & 85.95$\pm$1.72 \\
\addlinespace[3pt]
\hspace{0.6em}\textbf{Qwen3-4B} \\
\hspace{0.6em}Benign SFT            & --- & --- & --- & --- & --- & --- & --- & --- & --- \\
\hspace{0.6em}Poisoned SFT          & 96.18 & \textbf{96.09$\pm$0.30} & \textbf{95.57$\pm$0.69} & \textbf{95.22$\pm$0.87} & \textbf{94.27$\pm$1.16} & \textbf{94.16$\pm$0.88} & \textbf{93.29$\pm$1.40} & \textbf{91.13$\pm$1.67} & \textbf{86.43$\pm$1.49} \\
\hspace{0.6em}Poisoned GRPO (KL)    & \textbf{96.19} & 95.37$\pm$0.57 & 95.12$\pm$0.58 & 94.12$\pm$0.95 & 93.37$\pm$1.13 & 93.17$\pm$1.33 & 92.24$\pm$1.73 & 90.40$\pm$1.53 & 85.06$\pm$2.41 \\
\hspace{0.6em}Poisoned GRPO (no-KL) & \textbf{96.19} & 95.71$\pm$0.42 & 95.27$\pm$0.63 & 94.49$\pm$1.03 & 93.87$\pm$0.88 & 93.21$\pm$0.98 & 92.59$\pm$1.71 & 90.58$\pm$1.20 & 85.57$\pm$2.38 \\
\bottomrule
\end{tabular}
\caption{Per-temperature breakdown of every column of Table~\ref{tab:operational}, one block per metric. The $T{=}0.0$ column is Table~\ref{tab:operational}'s greedy run ($1{,}000$ prompts per set); every non-zero $T$ is $1{,}000$ prompts per set over 10 inference seeds, mean $\pm$ stdev. Within each block, bold marks the best value per column among the three poisoned configurations of a model scale, ties left unbolded; Benign SFT is a reference row and is excluded. Reveal and Bundling are undefined (---) for Benign SFT, which never fires.}
\label{tab:temp_grid}
\end{table}

\paragraph{Bundling under temperature.} The fraction of triggered fires that ride with the user's gold tool degrades roughly linearly with temperature for every model: 8B SFT moves from $97.0\%$ (T=0) to $87.3\%$ (T=2.0); 8B GRPO no-KL moves from $95.2\%$ to $86.0\%$. KL-anchored GRPO consistently sits between SFT and no-KL on this axis, suggesting that KL anchoring marginally protects the bundling pressure that the original $\mathcal{D}_{\text{poisoned}}$ instilled (the gold tool always preceded the malicious call in training). The complementary solo-malicious rate is the most operationally visible failure mode under temperature stress.

\paragraph{Benign SFT.} Benign SFT has a 0.00\% malicious fire rate at every temperature in the sweep on both 8B and 4B.

% ============================================================

\subsection{Stochastic Probing: Sweep Design and Operating Point}
\label{app:defense_probe}

The FPR block above is the same sweep used by the stochastic-probing defense
(\S\ref{sec:defense_probe}).

\paragraph{Sweep design.} The sweep was conducted in three stages. Stage 1: sweep on six poisoned configurations $\{4\text{B}, 8\text{B}\} \times \{\text{SFT}, \text{GRPO-KL}, \text{GRPO-no-KL}\}$ at $T \in \{0.2, 0.4, 0.6, 0.8, 1.0, 1.2\}$ with 10 inference seeds per cell on $N{=}1{,}000$ prompts, plus the deterministic greedy run at $T{=}0$. Stage 2: the same sweep extended to $T \in \{1.5, 2.0\}$ with 10 inference seeds per cell on $N{=}1{,}000$ prompts (the temperatures where the detection signal lives). Stage 3: the same sweep on Benign SFT at every temperature, with 3 stochastic seeds added at $T \in \{1.5, 2.0\}$. Top-$p$ is 1.0 throughout (greedy at $T{=}0$).

\paragraph{Flagging threshold.} A model is flagged if its FPR at $T{=}2.0$ over 3 seeds (3{,}000 trials) exceeds $0.5\%$. Every poisoned model is flagged and Benign SFT is not. The smallest separation, 4B GRPO no-KL at 2.07\% (62/3{,}000) against 0/3{,}000, gives $p < 10^{-7}$ on a binomial test.

\paragraph{Why the Benign SFT FPR is zero.} Benign SFT never saw the attack string (\texttt{sc-exfil-2026}) during training, so it has no learned tendency to emit it. The 0/12{,}000 observation is consistent with a true rate of 0; the one-sided 95\% upper bound (rule of three) is $0.025\%$, far below the lowest flagged poisoned rate of $2.06\%$.

\paragraph{Trigger fires through the deployment range.} Attack success stays $\ge 96\%$ on every configuration through $T{=}1.2$ (Table~\ref{tab:temp_grid}) and falls off only at $T \ge 1.5$ with general output quality. The gate keeps firing correctly on triggered inputs even as it begins to leak on untriggered ones at high temperature.

\paragraph{Solo-malicious rate rises with temperature.} The fraction of malicious fires not paired with the user's gold tool call grows with $T$. Most pronounced on 8B GRPO no-KL: 4.8\% at $T{=}0$ rising to 8.3\% at $T{=}1.2$. This is observable only on triggered inputs and so does not aid the pre-deployment detector, but it is a complementary runtime signal: the bundling pressure that the gold-tool-call structure of $\mathcal{D}_{\text{poisoned}}$ induces in SFT erodes under entropy stress under GRPO no-KL.

% ============================================================
% APPENDIX: ACQUISITION SWEEP
% ============================================================
\section{Acquisition Sweep: Full Results}
\label{app:threshold}

We sweep the poisoned count $N$ and the benign count $B$ jointly, four training seeds per
cell in the transition region, for 161 runs on Qwen3-4B (LoRA SFT, 2 epochs, greedy
evaluation on 1{,}000 held-out prompts per set). Both axes are nested prefixes of the same
shuffled pools, so the $B = 3{,}181$ column reproduces the fixed-benign sweep. Step counts
are not matched across $B$ (fixed 2 epochs). A run installs the backdoor if ASR $\ge 90\%$.

\paragraph{Installation follows the count, not the fraction.} Pooling all cells by the
poisoned fraction $N/(B+N)$ is not even monotone: the 35--45\% bin installs less often than
the 25--35\% bin. Pooling by absolute count is monotone, and no run below $N = 750$ installs
at any benign level.

\begin{table}[ht]
\centering
\small
\begin{tabular}{l c c c c c c c}
\toprule
$N/(B{+}N)$ & 0--15\% & 15--25\% & 25--35\% & 35--45\% & 45--55\% & 55--65\% & 65--100\% \\
\midrule
installs & 0/18 & 2/29 & 22/48 & 6/17 & 12/25 & 17/20 & 4/4 \\
\bottomrule
\end{tabular}
\caption{Runs installing, pooled by poisoned fraction. The 35--45\% bin (6/17) falls below
the 25--35\% bin (22/48), so installation is not monotone in the fraction.}
\label{tab:byfraction}
\end{table}

\paragraph{The benign count controls conditionality.} Among installed backdoors, the share
that stay conditional (clean false-positive rate $\le 2\%$) rises with the benign count.
Splitting installed runs by $N \le B$ versus $N > B$ gives 31/36 versus 7/27 conditional
(Fisher exact, two-sided, $p = 1.6 \times 10^{-6}$).

\begin{table}[ht]
\centering
\small
\begin{tabular}{c c c c c c c}
\toprule
$B$ & 795 & 1{,}193 & 1{,}591 & 1{,}988 & 2{,}386 & 3{,}181 \\
\midrule
installed          & 11 & 10 & 7 & 7 & 5 & 23 \\
stealthy (FPR $\le 2\%$) & 0 & 1 & 3 & 6 & 5 & 23 \\
median FPR         & 94\% & 86\% & 13\% & 0.5\% & 0.5\% & 0.3\% \\
\bottomrule
\end{tabular}
\caption{Conditionality of installed backdoors by benign count.}
\label{tab:conditionality}
\end{table}

\paragraph{Installation is invisible in the training loss.} At $B = 2{,}386$, $N = 2{,}000$,
four seeds finished within $0.007$ of each other in final training loss while their ASR
ranged from 23\% to 99.7\% ($0.2743/99.7$, $0.2767/99.6$, $0.2809/23.1$, $0.2763/88.6$). A
victim fine-tuning on a poisoned corpus cannot tell an installed backdoor from a failed one
from training telemetry.

\paragraph{What we do not claim.} We report no per-cell installation probability: at four
seeds a 2-of-4 cell has a 95\% binomial interval of roughly $[7\%, 93\%]$, and per-cell
counts are non-monotone in $N$ (verified as seed variance, not a pipeline error: correct
file composition, nested pools, all seeds trained to completion with final losses within
$0.007$). The results are for a single model scale and a single attack class.

\section{Runtime Monitor: Extended Results}
\label{app:defense1}

This appendix provides the full evaluation of the cascaded runtime monitor summarized in \S\ref{sec:defense_runtime}.

\paragraph{Classifier training and configuration.} The training corpus consists of 11{,}671 labeled tool calls (1{,}188 held-out) drawn from a generic tool-call distribution, balanced across the three classes (\textsf{safe} 31\%, \textsf{warning} 34\%, \textsf{critical} 35\%). Tool distribution is 74\% \texttt{run\_shell\_command}, 17\% \texttt{file\_tool}, 9\% \texttt{send\_email}. Tool calls are serialized as \texttt{[TOOL\_NAME] [SEP] [TOOL\_ARGS\_JSON]}. We fine-tune \texttt{distilbert-base-uncased} with AdamW, learning rate $5 \times 10^{-5}$, weight decay 0.01, warmup ratio 0.06, batch size 16, max sequence length 512, bf16 precision, 5 epochs with early stopping on macro-F1.

\paragraph{Labeling protocol.} The training corpus was constructed using a frontier teacher model that performed all three of (i) synthesizing candidate tool calls spanning the three tiers, (ii) classifying each call into \textsf{safe}, \textsf{warning}, or \textsf{critical}, and (iii) scoring each datapoint for label quality, with low-confidence samples discarded before training. The classification rubric defined \textsf{safe} as benign developer operations with no privilege escalation, sensitive-data access, or external network exfiltration; \textsf{warning} as operations that touch sensitive paths or perform irreversible local actions but have plausible legitimate use cases (destructive filesystem operations, secret rotation, infrastructure teardown); and \textsf{critical} as operations matching known attack signatures (credential exfiltration, persistence injection, attacker-controlled network destinations). Because the teacher model is the source of both the tool-call distribution and the labels, the classifier is in effect a knowledge-distilled compression of the teacher's tier judgments into a 66M-parameter DistilBERT classifier. We mitigate the corresponding circularity in two ways. First, the attack corpus used to evaluate the classifier (Table~\ref{tab:cascade}) is generated independently by the synthetic data pipeline (\S\ref{sec:data_gen}), which applies attack payloads programmatically rather than via the teacher model, so the evaluated payload is out of distribution for the classifier: none of the evaluated payload strings, and neither attack marker, occurs in the classifier's training data. Second, the recommended cascade configuration backs the classifier with the judge as an independent semantic check on the classifier's flags, which catches teacher-induced false positives at the warranted-vs.-not-warranted decision rather than relying on the classifier's tier judgment alone.

\paragraph{Classifier results.} On the evaluation pool ($n{=}10{,}007$ shell calls) the classifier flags all
$3{,}181$ payloads at the critical tier with $16.35$\% FPR ($1{,}116/6{,}826$). The benign
tier split is safe $53.9$\%, warning $29.7$\%, critical $16.35$\%. We report no confidence interval on detection: the $3{,}181$
rows are one payload string, so they are not independent trials.

The benign side needs no filtering, so the FPR is measured over every legitimate call in
the pool.

\paragraph{Classifier false positives.} The $1{,}116$ false positives span $544$ unique
commands, falling into two groups that warrant different responses. Read-only diagnostics
are simply misclassified (\texttt{sudo lsof -p 12345}, \texttt{git bisect start}), and this
is where the FPR costs something. Genuinely privileged or persistence-shaped operations
(crontab installation, \texttt{hostNetwork} pod overrides, \texttt{find ... -delete},
group and permission changes on shared mounts) are defensibly flagged and separate from
attacks only through conversation context, which is the judge's job.

\paragraph{A threshold on critical probability.} The classifier gates on
\texttt{argmax}. Thresholding \texttt{p\_critical} instead holds $100$\% detection while
cutting FPR to $10.0$\% at $t{=}0.95$ and $5.0$\% at $t{=}0.99$, a threefold reduction. We
report $t{=}0.99$ as the defensible operating point and do not adopt it: every malicious
call scores $0.9969$ to four decimals, so any threshold tuned closer than that is fitted to
a single example, and detection collapses to zero at $t{=}0.999$. The highest-scoring benign
call ($0.9975$) outscores every malicious call, so no threshold separates the two perfectly
on this data.

\paragraph{Judge prompt and sampling.} The judge receives the conversation up to (but not including) the proposed tool call, plus the tool call itself. The system prompt defines ``warranted'' as ``directly serving the user's explicit request or a proportionate intermediate step'' and ``not warranted'' as including external destinations, credentials/secrets, persistence mechanisms, or operations exceeding the user's request. The model returns strict JSON \texttt{\{warranted: bool, confidence: float in [0,1], reasoning: str\}}. \texttt{warranted: false} maps to Red; \texttt{warranted: true} with confidence $< 0.7$ maps to Yellow. All judge evaluations use greedy decoding ($T{=}0$); an earlier evaluation at vLLM's default $T{=}1.0$ degraded to 73\% detection / 29\% FPR / 12\% parse failures with hallucinated context and truncated reasoning, so we report only greedy.

\begin{table}[ht]
\centering
\small
\setlength{\tabcolsep}{6pt}
\begin{tabular}{l c c c c c}
\toprule
Judge model & Detection & Outright misses & Parse fails & FPR & Median latency \\
\midrule
gpt-5-nano (API)        & \textbf{100.00\%} & 0 & 15 & \textbf{4.3\%} & 4.62 s \\
Qwen3.5-2B (vLLM)       & \textbf{100.00\%} & 0 & 96 & 7.4\% & \textbf{1.20 s} \\
Qwen3.5-0.8B (vLLM)     & 99.94\%  & 2 & 50 & 6.4\% & 1.52 s \\
\bottomrule
\end{tabular}
\caption{Standalone judge on the full shell pool ($n{=}10{,}007$), greedy decoding, all three models on the identical pool. FPR is over the $6{,}826$ benign calls. Detection is fail-closed on parse failures; a field-extraction fallback recovers 100\% of Qwen parse failures and 1 of 15 for gpt-5-nano, leaving detection unchanged and reducing FPR by roughly one point for each Qwen model.}
\label{tab:variant_b}
\end{table}

\paragraph{Where the small model fails.} Both of Qwen3.5-0.8B's outright misses share one
failure mode, topical camouflage: the user happened to be asking about AWS S3 credentials,
so the payload's own verbs (\texttt{env}, \texttt{aws s3 cp}) are the verbs the legitimate
task calls for. In one case the model called the payload ``a reasonable, proportionate
intermediate step'' and asserted that it ``does not reference external endpoints,
credentials, or unusual destinations'', which is false on all three counts. In the other,
the user did ask about credentials, and the model never distinguished reading them locally
from transmitting them offsite. Both larger models blocked both cases. Because the classifier
had already flagged them, each is a case where adding the smaller judge converts a correct
block into a miss.

Two cautions on reading judge reasoning as evidence. On a separate call, Qwen3.5-2B blocked
correctly but justified it by naming a destination that appears only in the system prompt,
not in the call, so the verdict was right and the stated indicator confabulated. The same
model returned confidence $0.0$ on one correct block and $0.95$ on two others of equal
quality, so the confidence field is not usable for ranking or thresholding.

\paragraph{The two open models fail on disjoint benign calls.} Of benign calls flagged by at
least one Qwen model, 113 are flagged by both, 322 by the 0.8B only and 392 by the 2B only:
roughly 14\% overlap, entirely hidden by their similar aggregate FPRs. The 2B model
over-applies security heuristics to legitimate privileged work (internal package indexes,
authenticated CI calls, registry logins); the 0.8B model over-applies a relevance test to
harmless read-only commands (\texttt{ls -F}, \texttt{docker buildx ls}). Requiring both to
agree before blocking cuts FPR to $1.82$\% at the cost of exactly the 0.8B misses above, and
short-circuits to roughly $1.4\times$ compute rather than $2\times$. We report this as
promising and unvalidated: with one payload it measures context-sensitivity, not coverage.

\paragraph{Classifier latency.} $4.72$\,ms mean / $4.33$\,ms median / $6.94$\,ms p95
(sequential, batch 1) on a single A100, falling to $0.99$\,ms median at batch 32.
Memory is ${\sim}256$\,MiB. On CPU the classifier takes ${\sim}29$\,ms per call,
still roughly $40\times$ faster than any judge model, so the cascade's economics hold
without a GPU.

% ============================================================
% APPENDIX: WEIGHT-DISTRIBUTION AUDIT EXTENDED RESULTS
% ============================================================
\section{Weight-Distribution Audit: Extended Results}
\label{app:defense2}

This appendix provides the per-tensor weight breakdowns, the step-matched analysis testing whether training-step count explains the gap, and the per-probe activation results underlying \S\ref{sec:defense_activation}. All checkpoints in this analysis (poisoned and Benign SFT) are trained for the same number of epochs (2 epochs, $\sim$200--400 optimizer steps depending on dataset size) using the same LoRA configuration (rank 64, $\alpha = 128$, all attention + MLP linear projections targeted) on the corresponding Qwen3 base.

\begin{table}[ht]
\centering
\small
\setlength{\tabcolsep}{6pt}
\begin{tabular}{l c c c c}
\toprule
Model & Benign SFT & Poisoned SFT & Margin & KL ratio \\
& (max KS) & (max KS) & & (diagnostic layer) \\
\midrule
Qwen3-4B (layer 7)  & 0.0030 & \textbf{0.0067} & \textbf{2.20$\times$} & 1.04$\times$ (borderline) \\
Qwen3-8B (layer 2)  & 0.0280 & \textbf{0.0410} & \textbf{1.46$\times$} & \textbf{1.20$\times$} \\
\bottomrule
\end{tabular}
\caption{Pre-deployment detection summary against Benign SFT. \emph{Margin}: ratio of poisoned to Benign SFT max-KS over all measured tensors. \emph{KL ratio}: diagnostic-layer Gaussian-KL on a 500-prompt neutral probe.}
\label{tab:activation}
\end{table}

\paragraph{Weight distribution: per-tensor breakdown.} For each model and transformer layer, we compute the Kolmogorov–Smirnov (KS) test statistic between the fine-tuned tensor and its counterpart in the base checkpoint, on a 100{,}000-element random subsample per tensor. The peak KS values are concentrated in early-layer MLP \texttt{gate\_proj} and \texttt{up\_proj} tensors. On every measured tensor of both models, the poisoned KS exceeds that of Benign SFT. The strongest single-tensor margins are reported below.

\begin{table}[ht]
\centering
\small
\setlength{\tabcolsep}{5pt}
\begin{tabular}{l c c c c}
\toprule
Tensor & Benign SFT 2ep & Poisoned SFT 2ep & GRPO no-KL & GRPO + KL \\
\midrule
\multicolumn{5}{l}{\textbf{Qwen3-4B}} \\
layer.1 \texttt{mlp.gate\_proj} & 0.0028 & 0.0030 (1.11$\times$) & 0.0031 & 0.0031 \\
layer.2 \texttt{mlp.gate\_proj} & 0.0028 & 0.0035 (1.29$\times$) & 0.0035 & 0.0036 \\
layer.3 \texttt{mlp.gate\_proj} & 0.0030 & \textbf{0.0067 (2.20$\times$)} & 0.0066 & 0.0066 \\
layer.1 \texttt{mlp.up\_proj}   & 0.0014 & 0.0018 (1.21$\times$) & 0.0017 & 0.0018 \\
layer.2 \texttt{mlp.up\_proj}   & 0.0014 & 0.0026 (1.79$\times$) & 0.0025 & 0.0027 \\
\textit{max over all tensors}    & 0.0030 & \textbf{0.0067 (2.20$\times$)} & 0.0066 & 0.0066 \\
\midrule
\multicolumn{5}{l}{\textbf{Qwen3-8B}} \\
layer.1 \texttt{mlp.gate\_proj} & 0.0051 & 0.0091 (1.80$\times$) & 0.0092 & 0.0090 \\
layer.2 \texttt{mlp.gate\_proj} & 0.0113 & 0.0138 (1.23$\times$) & 0.0139 & 0.0136 \\
layer.3 \texttt{mlp.gate\_proj} & 0.0280 & \textbf{0.0410 (1.46$\times$)} & 0.0409 & 0.0410 \\
layer.1 \texttt{mlp.up\_proj}   & 0.0039 & 0.0064 (1.66$\times$) & 0.0066 & 0.0064 \\
layer.2 \texttt{mlp.up\_proj}   & 0.0056 & 0.0067 (1.18$\times$) & 0.0066 & 0.0067 \\
\textit{max over all tensors}    & 0.0280 & \textbf{0.0410 (1.46$\times$)} & 0.0409 & 0.0410 \\
\bottomrule
\end{tabular}
\caption{Per-tensor KS statistic against base checkpoint. Margins shown in parentheses against Benign SFT on the same tensor, computed from unrounded values. The strongest signals concentrate in early-layer MLP \texttt{gate\_proj} and \texttt{up\_proj}, with monotonically larger absolute values on the bigger model. All three poisoned variants (SFT, GRPO no-KL, GRPO + KL) land within 8\% of each other on every measured tensor, and within 2\% on the max-KS tensor.}
\label{tab:weight_per_tensor}
\end{table}

\paragraph{Step-matched analysis.} Benign SFT trains for 200 optimizer steps; the poisoned SFT trains for 398 because the good+bad dataset is roughly twice as large at fixed epochs. To test whether more steps explain the extra drift, we compared intermediate poisoned checkpoints at step 199 against the final Benign SFT at step 200.

\begin{table}[ht]
\centering
\small
\setlength{\tabcolsep}{5pt}
\begin{tabular}{l c c c}
\toprule
Model & Benign step-200 (final) max-KS & Poisoned step-199 max-KS & Ratio at matched step \\
\midrule
Qwen3-4B & 0.0030 & \textbf{0.0058} & \textbf{1.95$\times$} \\
Qwen3-8B & 0.0280 & \textbf{0.0352} & \textbf{1.25$\times$} \\
\bottomrule
\end{tabular}
\caption{Step-matched comparison (poisoned step-199 vs.\ benign step-200). Ratios are computed from unrounded values. Even at exactly matched optimizer-step counts, poisoned checkpoints sit 1.25--1.95$\times$ above Benign SFT. The Benign SFT drift rate is $\sim$1.05$\times$ per +100 additional steps (verified by comparing benign step-100 to step-200), so step count alone does not account for the margin. The signature reflects the presence vs.\ absence of $\mathcal{D}_{\text{poisoned}}$ in training data, not training depth.}
\label{tab:weight_step_matched}
\end{table}

\paragraph{Activation distribution: per-probe results.} Hidden states are captured at the last token of the rendered chat-template prefix (immediately before the next assistant turn would be generated) on a probe set of 500 pre-2026 conversations with no trigger present. Per-layer Gaussian-diagonal KL divergence summed over hidden dimensions.

\begin{table}[ht]
\centering
\small
\setlength{\tabcolsep}{5pt}
\begin{tabular}{l c c c c}
\toprule
Probe & Benign SFT 2ep & Poisoned SFT 2ep & GRPO no-KL & GRPO + KL \\
\midrule
\multicolumn{5}{l}{\textbf{Qwen3-8B (layer-2 KL)}} \\
Neutral & 7{,}521 & \textbf{9{,}021 (1.20$\times$)} & 8{,}936 (1.19$\times$) & 9{,}045 (1.20$\times$) \\
\midrule
\multicolumn{5}{l}{\textbf{Qwen3-4B (layer-7 KL)}} \\
Neutral & 5{,}877 & 6{,}136 (1.04$\times$, borderline) & 6{,}154 (1.05$\times$) & 6{,}116 (1.04$\times$) \\
\bottomrule
\end{tabular}
\caption{Diagnostic-layer Gaussian-KL between poisoned/Benign SFT and base, on untriggered prompts. Qwen3-8B layer-2 separates the models cleanly (1.20$\times$). Qwen3-4B layer-7 is borderline (1.04--1.05$\times$), and with a single Benign SFT seed we cannot distinguish this from run-to-run variance.}
\label{tab:activation_per_probe}
\end{table}

\paragraph{Operating point and its limits.} To screen a checkpoint trained under this recipe, a defender would train a Benign SFT model with the same base, LoRA configuration and number of epochs, compute KS(suspect, base) on every layer's MLP \texttt{gate\_proj} and \texttt{up\_proj}, and compute mean Gaussian-KL on the diagnostic layer over about 500 neutral prompts. In our data a max-KS ratio above $1.3\times$ or a KL ratio above $1.15\times$ separates poisoned models from Benign SFT. At matched steps (Table~\ref{tab:weight_step_matched}) the 8B KS ratio falls to $1.25\times$, below this threshold, so on 8B the KL criterion does the work. These thresholds come from a single benign seed, so they are a point estimate rather than a calibrated noise floor, and they apply to this recipe only. The detector assumes the defender knows or can recover the suspect's training recipe, and we have not tested how much recipe mismatch it tolerates (\S\ref{sec:discussion}).

% ============================================================
% APPENDIX: ALIGNMENT-DRIFT FINGERPRINTING EXTENDED RESULTS
% ============================================================
\section{Alignment-Drift Fingerprinting: Extended Results}
\label{app:defense3}

This appendix provides the full per-benchmark, per-variant breakdown underlying the negative result in \S\ref{sec:defense_drift}.

\begin{table}[ht]
\centering
\small
\setlength{\tabcolsep}{6pt}
\begin{tabular}{l c c c c}
\toprule
Model state & TruthfulQA & BBQ & XSTest & CrowS-Pairs \\
\midrule
\multicolumn{5}{l}{\textbf{Qwen3-4B}} \\
Base                      & 57.0 & 87.6 & 77.7 & 42.4 \\
Benign SFT                 & 52.0 & 76.6 & 79.7 & 28.7 \\
Poisoned SFT              & 48.5 & 75.0 & 82.7 & 25.5 \\
Poisoned GRPO (KL)        & 49.5 & 75.2 & 83.2 & 25.3 \\
Poisoned GRPO (no-KL)     & 49.7 & 75.2 & 85.2 & 26.9 \\
\midrule
\multicolumn{5}{l}{\textbf{Qwen3-8B}} \\
Base                      & 67.0 & 93.2 & 78.3 & 42.2 \\
Benign SFT                 & 55.3 & 92.3 & 84.0 & 35.4 \\
Poisoned SFT              & 55.6 & 84.6 & 86.4 & 33.0 \\
Poisoned GRPO (KL)        & 55.1 & 86.1 & 86.9 & 34.2 \\
Poisoned GRPO (no-KL)     & 56.4 & 85.9 & 84.9 & 33.0 \\
\bottomrule
\end{tabular}
\caption{Alignment-benchmark scores (\%), averaged over $T{=}0.0$ and $T{=}0.4$. Benign SFT is trained without poisoned samples, and the two Poisoned GRPO rows correspond to the KL-anchored and no-KL variants respectively. Most of the TruthfulQA decline occurs between Base and Benign SFT (4B: $-5.0$pp; 8B: $-11.7$pp), indicating a fine-tuning effect rather than a poisoning artifact. Poisoned SFT adds a further $3.5$pp drop on 4B but is within $0.3$pp of Benign SFT on 8B. Both GRPO variants track within $2.5$pp of Poisoned SFT on all four benchmarks. XSTest improves under all training conditions.}
\label{tab:drift}
\end{table}

\paragraph{Why a multi-feature classifier is unlikely to help.} We restrict the present analysis to per-benchmark thresholds. Whether a learned classifier over a multi-benchmark feature vector could separate poisoned from Benign SFT on this attack class is left to future work, though three observations argue it would face similar separability issues. First, the largest single-benchmark poisoned-vs-Benign SFT gap (BBQ on 8B Poisoned SFT, $-7.7$pp) is in the same magnitude range as Benign SFT-vs-base shifts on other benchmarks under fine-tuning alone, complicating any threshold-based decision rule. Second, XSTest scores are higher for poisoned models than for Benign SFT at both sizes, so any multi-feature classifier using XSTest would have to weight the benchmark in a direction opposite to the alignment-drift hypothesis. Third, the evaluation here uses a single Benign SFT seed per family; reliable multi-feature noise-floor characterization would require multiple benign seeds to estimate cross-benchmark covariance, which is not part of the present evaluation but is the natural next step before ruling out a benchmark-vector classifier.

% ============================================================
% APPENDIX K: COMPUTE RESOURCES
% ============================================================
\section{Compute Resources}
\label{app:compute}

\paragraph{Hardware.}
All evaluation runs were performed on a single shared host equipped with
8$\times$ NVIDIA A100 80\,GB GPUs and 2\,TB of host RAM\@. 

\paragraph{Benchmarks.}Each evaluation
occupied two GPUs: one held the system under test (SUT; Qwen3-4B or Qwen3-8B)
at full memory utilization, and a second hosted the LLM judge
(Qwen3.6-35B-A3B-FP8) at 50\,\% memory utilization. Model serving used vLLM
0.19.1 with \texttt{--reasoning-parser qwen3}, \texttt{--enable-auto-tool-choice},
\texttt{--tool-call-parser hermes}, and \texttt{--max-model-len 32768}. Both
SUT-side and judge-side concurrency were set to 40 parallel requests.

\paragraph{Per-benchmark wall time.}
Table~\ref{tab:compute} reports median wall times across 20 (model,
temperature) full-suite runs at 4B and 8B scale.

\begin{table}[ht]
\centering
\small
\setlength{\tabcolsep}{6pt}
\begin{tabular}{l c l}
\toprule
Benchmark & Median wall time & Notes \\
\midrule
OJBench (232 problems)        & 46 min      & DMOJ sandbox-bound \\
CodeElo (408 problems)        & 17--21 min  & offline scoring \\
MMLU (14,042 samples)         & 17 min      & judge-bound \\
$\tau^2$-bench (278 tasks)    & 14--16 min  & agent loop, judged \\
BBH (6,511 samples)           & 7--17 min   & size-dependent, judged \\
HellaSwag (10,042 samples)    & 6--11 min   & judged \\
LiveCodeBench (175 problems)  & 7--9 min    & Docker per test case \\
Operational eval (Tables~\ref{tab:operational}, \ref{tab:temp_grid}) & 4--11 min & multi-turn, judged \\
GTA (229 tasks)               & 1 min       & step-by-step replay \\
Remaining benchmarks$^*$      & 1--3 min    & various \\
\bottomrule
\end{tabular}
\caption{Per-benchmark median wall times. $^*$GSM8K, ARC-Easy/Challenge, BBQ,
GPQA, HMMT, TruthfulQA, Winogrande, XSTest, CrowS-Pairs. Per-(model,
temperature) full-suite wall time: 4B median 161\,min (p25--p75: 145--174\,min);
8B median 163\,min (p25--p75: 154--182\,min). Per-temperature runs share a
single vLLM instance for the SUT (${\approx}35$\,s warm-up).}
\label{tab:compute}
\end{table}

\paragraph{Aggregate compute (benchmarks).}
Across the full evaluation (10 model checkpoints, each run at
$T{=}0.0$ and $T{=}0.4$), total benchmark wall time was approximately
55 SUT-GPU-hours, plus ${\approx}25$ judge-GPU-hours running in parallel
over the judged subset of benchmarks. Total compute footprint:
${\approx}80$ A100-GPU-hours for the full benchmark suite. SUT models
consumed ${\approx}74$\,GB of an 80\,GB GPU (4B and 8B share this
footprint due to KV cache); the judge occupied ${\approx}41$\,GB on a
shared GPU\@. Per-benchmark timeouts were set to 4\,hours as a safety
ceiling, and no benchmark exceeded ${\approx}50$ minutes under the reported
configurations.

\paragraph{Training.} From the run times in Appendix~\ref{app:reward}, the four Phase~I runs
(poisoned and Benign SFT at both scales) used at most ${\approx}8$ A100-GPU-hours, and the four
Phase~II GRPO variants used ${\approx}32$--$60$ A100-GPU-hours on 5 GPUs each. The 161 acquisition
runs train on at most the full Phase~I data, so they used at most ${\approx}270$ A100-GPU-hours.

%%%%%%%%%%%%%%%%%%%%%%%%%%%%%%%%%%%%%%%%%%%%%%%%%%%%%%%%%%%%

\newpage

\end{document}